\documentclass[a4paper,11pt]{article}
\pdfoutput=1

\usepackage{jheppub}
\usepackage[T1]{fontenc}
\usepackage{tensor}
\usepackage{subcaption}
\hypersetup{bookmarksnumbered}
\usepackage{bbold}
\usepackage{float}

\usepackage{lipsum}

\newcommand{\be}{\begin{equation}}
\newcommand{\ee}{\end{equation}}

\newcommand{\bea}{\begin{eqnarray}}
\newcommand{\eea}{\end{eqnarray}}

\newcommand{\ba}{\begin{equation}\begin{aligned}}
\newcommand{\ea}{\end{aligned}\end{equation}}

\def\legP{\mathcal{P}}
\def\avg#1{\Big\langle#1\Big\rangle}
\def\tr#1{\text{Tr}\left(#1\right)}

\usepackage[dvipsnames]{xcolor}

\title{Bootstrapping mesons at large N\\ 
{\Large Regge trajectory from spin-two maximization}}

\author{Jan Albert$^{\pi,\rho}$,}
\author{Johan Henriksson$^{f_2,\rho_3}$,}
\author{Leonardo Rastelli$^{\pi}$,}
\author{and Alessandro Vichi$^{f_2}$}
\affiliation{$^{\pi}$C. N. Yang Institute for Theoretical Physics, Stony Brook University,\newline Stony Brook, NY 11794-3840, U.S.A.}
\affiliation{$^{\rho}$Simons Center for Geometry and Physics, Stony Brook University,\newline Stony Brook, NY 11794-3636, U.S.A.}
\affiliation{$^{f_2}$ Department of Physics, University of Pisa and INFN, Largo Pontecorvo 3, I-56127 Pisa, Italy}
\affiliation{$^{\rho_3}$ Universit\'e Paris--Saclay, CEA, Institut de Physique Th\'eorique, 91191, Gif-sur-Yvette, France}

\preprint{YITP-SB-2023-41}

\abstract{We continue the investigation of large $N$ QCD from a modern bootstrap perspective, focusing on the mesons.
We make the natural spectral assumption 
that the  $2 \to 2$ pion amplitude  must contain,
above the spin-one rho meson,
a massive resonance of spin two. By maximizing its coupling we find 
a very interesting extremal  solution of the dual bootstrap problem, 
which
appears to contain at least a full Regge trajectory. Its low-lying states are in uncanny quantitative agreement with the  meson masses in the real world.

}

\begin{document} 
\maketitle

\section{Introduction}

\begin{figure}[h]
\centering
\includegraphics[width=0.75\textwidth]{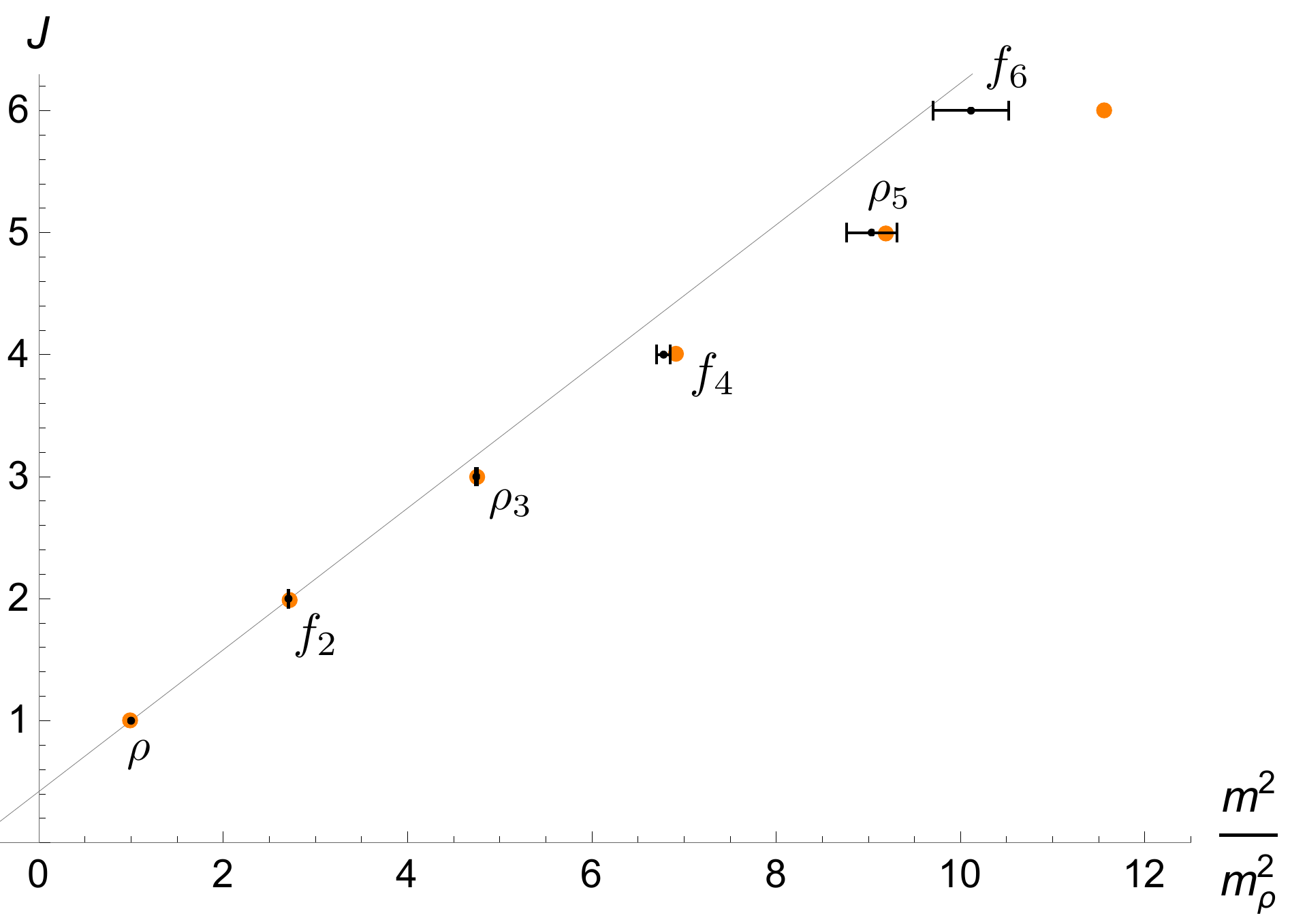}
\caption{Our best estimate for the extremal spectrum at the $f_2$ kink (orange points), together with the spectrum of real-world mesons as listed in \cite{PDG} (with error bars). To guide the eye, we have drawn a linear trajectory passing through the $\rho$ and $f_2$ 
mesons (gray line).}
    \label{fig:physical-spectrum}
\end{figure}

This article extends our exploration of large $N$ QCD by modern bootstrap methods. We focus on the mesons, particularly on the $2 \to 2$ pion amplitude.  We make the natural spectral assumption that a massive resonance of spin two  must appear
as intermediate state above the spin-one rho meson.
 Through maximization of its coupling we uncover a very intriguing extremal solution to the dual bootstrap problem. This solution seemingly contains a complete Regge trajectory and exhibits low-lying states that agree surprisingly well with the meson masses observed in the real world (figure \ref{fig:physical-spectrum}).

\subsubsection*{Recap}

The physical picture of large $N$ QCD  has long been clear. 
At strictly infinite $N$, its single-particle spectrum consists of  an infinite tower of stable, freely propagating  mesons and glueballs. To leading $1/N$ order,  these
asymptotic states interact via
 meromorphic scattering amplitudes, 
 with well-understood 
 high energy behavior. 
 This picture calls for the development of a bootstrap program that is in many ways parallel to the very powerful conformal bootstrap. We should consider the full landscape of putative large~$N$ confining gauge theories, 
 and rigorously carve it out  by imposing  physical consistency conditions on $2 \to 2$ scattering processes. The aspiration is that with enough physical input (such as suitable spectral assumptions) we will be able to {\it corner} large $N$ QCD at a special point in theory space.

A systematic investigation of this large $N$ theory space was initiated in \cite{Albert:2022oes} and further developed in \cite{Fernandez:2022kzi, Albert:2023jtd, Ma:2023vgc, Li:2023qzs}, focusing on the mesons.  Mesons form a consistent subsector at large~$N$ and are a natural place to start, both because their scattering is more constrained than that of glueballs
(due to flavor ordering of the external $q \bar q$ states) and because our explorations can be guided by the enormous wealth of real-world data.\footnote{
While in actual QCD we of course have $N=3$, it has long been appreciated that for many purposes $N=\infty$ is a surprisingly good approximation. Needless to say, our primary interest in the large $N$ theory goes beyond phenomenological considerations and it is ultimately driven  by the dream of finding the Platonic planar theory, which might have a dual string theory description.}  
The most obvious way to  parametrize theory space is in terms
of the {\it spectrum} of the full tower of large  $N$ mesons 
and all their on-shell {\it three-point couplings} (which are of order 
$O(1/\sqrt{N})$).\footnote{This is a mild oversimplification. Getting a bit ahead of our narrative,  three-point couplings would suffice if the Regge behavior allowed for {\it unsubtracted} dispersion relations. The need to make {\it one} subtraction means that a few four-point couplings are also needed to fully characterize all $2 \to 2$ scattering processes.}
These data are subject to the constraints of unitarity and crossing, in rather direct analogy with the conformal bootstrap.
A basic piece of spectral information comes from chiral symmetry breaking, which implies 
(if the quarks are massless, as we shall assume) the existence of massless Goldstone bosons, the pions, in the adjoint representation of
the $U(N_f)$ flavor group. There is then another, less direct but very useful parametrization of theory space.
Integrating out the massive mesons at tree level (as loops would be further suppressed in the $1/N$ expansion) one
obtains the large $N$ 
pion effective field theory (EFT),
 {\rm i.e.}\ the familiar chiral Lagrangian, and we can take
its infinite set of Wilson coefficients 
as specifying a point in theory space.
The cutoff $M$ of the EFT is naturally identified with the mass of the rho vector meson, the first exchanged massive state in pion scattering. 
One can then systematically enrich the analysis by progressively raising the cutoff, such that first rho and then the first few higher exchanged mesons are also included in the low-energy EFT.

In this framework, we are squarely within the program of constraining low-energy EFTs from UV consistency conditions, an old idea (see {\rm e.g.}\ \cite{Pham:1985cr, Pennington:1994kc,  Ananthanarayan:1994hf, Comellas:1995hq, Dita:1998mh, Adams:2006sv}) that has however been fully fleshed out only in recent years \cite{ Arkani-Hamed:2020blm, Bellazzini:2020cot, Tolley:2020gtv, Caron-Huot:2020cmc}.  
Fundamental properties of scattering amplitudes, such as unitarity, causality, crossing symmetry, the existence of a partial wave decomposition and Regge boundedness at high energy imply positivity bounds on the low-energy parameters.
We are in fact in the ideal scenario. Because our large $N$ EFT
is arbitrarily weakly coupled, 
the methods of~\cite{Caron-Huot:2020cmc} allow to derive rigorous two-sided bounds on homogeneous ratios of Wilson coefficients, rendered dimensionless by appropriate powers of the cutoff. 
The basic strategy is to write  dispersion relations, which relate the UV with the IR. One can systematically derive sum rules for the IR Wilson coefficients in terms of the UV spectral data, as well as ``null constraints'' (encoding crossing) that must be satisfied by the UV spectral data. The feasibility of these sum rules can be then recast into a convex optimization problem and solved with similar techniques extensively used in conformal bootstrap \cite{Rattazzi:2008pe,Poland:2011ey} (see \cite{Poland:2018epd,Rychkov:2023wsd} for technical reviews).

Even in the simplest setup one includes only the  pions in the low-energy EFT, this approach leads to surprisingly stringent constraints. 
Homogeneous ratios of Wilson coefficients (in units of the cutoff) are found to lie in compact regions whose size is of order one. The exclusion boundary in the two-dimensional space of four-derivative couplings 
displays an interesting geometry, with sharp corners and a tantalizing kink.  Injecting more physical assumptions, such as the presence of the spin-one rho meson, restricts the allowed space of EFT parameters, zooming in the region of interest. Some corners and edges of the allowed region were identified with simple scattering amplitudes~\cite{Caron-Huot:2020cmc}\cite{Albert:2022oes, Fernandez:2022kzi}, while some others remained unexplained. In particular, the straight segment that ends at the kink can be understood as a UV completion of a single rho exchange~\cite{Albert:2022oes, Fernandez:2022kzi}. 
All other known amplitudes, including stringy-like amplitudes such as the Lovelace--Shapiro's amplitude \cite{Lovelace:1968kjy,Shapiro:1969km}  live safely in the bulk of the allowed region.\footnote{Curiously, a version of the LS amplitude where the scalars have been subtracted appears to be very close to one of the exclusion boundaries, but still strictly inside the allowed region~\cite{Fernandez:2022kzi}.} Real world experimental data are also  compatible with the bounds, though their error bars are too large to draw any meaningful conclusion about where QCD sits.
 In addition, \cite{Albert:2023jtd, Ma:2023vgc} considered the EFT of massless pions coupled to background gauge fields,
a richer system that has access to a larger set of intermediate meson states and to the coefficient of the chiral anomaly.
 Compatibility of the dual and primal approaches to the pion EFT bounds was recently demonstrated in~\cite{EliasMiro:2022xaa,Li:2023qzs}.

Complementary to this line of work, the modern S-matrix bootstrap \cite{Paulos:2016fap,Paulos:2016but,Paulos:2017fhb} has developed systematic methods to construct the most general scattering amplitude consistent with the basic axioms of quantum field theory. By scanning over all possible amplitudes, one can explore the allowed values of several observables, such as interactions, masses of resonances, etc. Recent applications were also able to accommodate the low-energy behavior of an amplitude in order to reproduce a given EFT~\cite{Guerrieri:2021ivu,Karateev:2022jdb,Haring:2022sdp,Miro:2023bon,Guerrieri:2022sod}, while allowing the most general ultraviolet behavior. Leveraging  this approach, \cite{Guerrieri:2020bto,Guerrieri:2018uew,He:2023lyy, Guerrieri:2023qbg}  have revisited pion scattering and glueball in the non-perturbative, finite $N$ regime. An alternative interesting line of research focuses on the QCD flux tube~\cite{EliasMiro:2021nul,EliasMiro:2022xaa}.

\subsubsection*{Regge trajectories for pion scattering}

In this work 
we start from a simple observation: most of the explicit amplitudes saturating the bounds on Wilson coefficients either do not contain intermediate states with spin $J >1$, or if they do, they are clearly unphysical, with states of arbitrary high spin and equal mass, violating locality. 
On the other hand, QCD has a much richer spectrum, with resonances that organize themselves in Regge trajectories. In order to zoom on theories with similar properties we should inject some further physical input. The key assumption that we are going to make is the existence of a spin-two intermediate state in pion scattering. To understand the significance of this assumption, we need to recall some facts about the Regge limit.

It is a general fact about quantum field theory~\cite{Froissart:1961ux, Martin:1965jj} that
in the Regge limit of large Mandelstam~$s$
and fixed momentum transfer $u$,
scattering amplitudes are strictly bounded by $O(|s|^2)$.   Crucially for our story, the meson sector of large $N$ QCD is expected to have a softer Regge behavior. The pomeron Regge trajectory is suppressed at large $N$, and the leading trajectory is that of the spin-one massive rho meson, which has intercept $<1$.
 A large $N$ meson scattering amplitude must then satisfy 
\begin{equation}
\label{mesonRegge}
    \lim_{|s|\rightarrow\infty} \frac{M(s,u)}{s} = 0\, , \qquad \text{for fixed }u<0\,. 
\end{equation}
This behavior allows  to write dispersion relations with a single subtraction, while for general QFT 
amplitudes one would need  two subtractions. 

The exchange of a single spin-$J$ state leads to 
an amplitude that grows like $O(s^J)$ in the Regge limit. A $J=1$ exchange (such as the intermediate rho meson in pion scattering) does not satisfy our Regge assumption, but only  marginally. This is the intuitive reason  why one can ``UV complete it''
by adding an infinite set of higher spin states at a parametrically high scale $M_\infty \gg M$, whose purpose is to give the required softer Regge behavior without changing the low-energy Wilson coefficients, which are measured in units the cutoff $M$ \cite{Albert:2022oes,Fernandez:2022kzi}.
We expect however that the same mechanism won't work to UV complete an exchange with spin $J >1$, which strictly violates the assumed Regge bound -- a whole tower of arbitrarily high spins must conspire to give the desired suppression.\footnote{For weakly coupled gravitational theories,
the analogous statement  has been argued from thought experiments that leverage causality~\cite{Camanho:2014apa}. In that context, the marginal Regge behavior is $O(s^2)$, so ``higher spin'' must be interpreted as $J >2$.}
We make this intuition precise in section \ref{sec:theorems}, where we derive a series of spectral no-go theorems using null constraints.
Null constraints are identities that must be satisfied by the positive spectral density for it to be compatible with crossing and the assumed Regge behavior. We show that a single massive $J=2$ exchange cannot be fixed by adding states at arbitrary high scale: it forces the existence of at least one state with odd spin at a finite mass. The argument can be iterated: by choosing carefully the null constraints we can show that the existence of a single spin $J = 2$ requires the presence of additional states with larger and larger spin.\footnote{For concreteness, we have stated the version of these results
that applies directly to our current problem. Analogous theorems hold more generally. If an amplitude admits a $k-$subtracted dispersion relation, the presence of an intermediate state with $J >k$ will force a whole tower of higher spin states.}

This suggests an obvious strategy. We should enforce that
in addition to a massive $J=1$ state 
of mass $m_\rho$ (the rho meson), 
the pion amplitude must also contain a massive $J=2$ state of mass $m_{f_2} > m_\rho$ (the expected $f_2$ meson). The overall mass scale amounts to a choice of units, and for definiteness we tune the ratio $m_{f_2}/m_\rho$ to its real-world value. We leave the 
coupling $g_{\pi\pi f_2}$ of the $f_2$ to the external pions as a free parameter. For any non-zero $g_{\pi\pi f_2}$, we find that the amplitude is inconsistent unless additional states kick in at a finite value of the new cut-off $\widetilde M > m_{f_2}$. This is just what was expected from the no-go theorem described above. 
Rather wonderfully, the curve describing the maximum allowed value of $g_{\pi\pi f_2}$ as a function of $\widetilde M$ exhibits a sharp kink, which is numerically very stable, see figure~\ref{fig:kink-plot} below.
This kink (which we dub the ``$f_2$ kink'', to distinguish it from the old kink of~\cite{Albert:2022oes}) corresponds to a novel extremal solution of our  bootstrap problem.  

The discovery and numerical exploration of this extremal solution  are the principal results of this paper. 
A first striking fact is that ratio $g_{\pi\pi f_2}/g_{\pi\pi \rho}$
of the $f_2$ and rho on-shell couplings to the external pions is in perfect agreement with the real world value, see figure \ref{fig:couplings}.\footnote{
As we explain  below, the overall couplings (normalized by the pion decay constant), are somewhat smaller than in QCD, but this is just as expected. Our setup is insensitive to removing intermediate scalars and the best we can ever hope for is to zoom in on large $N$ QCD {\it with scalars subtracted}. Removing scalars would have precisely the desired effect of increasing the normalized couplings.}
Our extremal solution appears to contain a full Regge trajectory.
Figure \ref{fig:physical-spectrum}  shows our numerical estimates for its first few states,  together with the spectrum of the real-world mesons that appear in pion-pion scattering.
The agreement for $J=3, 4, 5$ is rather stunning
(recall that the  $J=1$ mass fixes the scale and the $J=2$ mass is an input). Our solution seems to accurately trace the small deviation from 
a linear Regge trajectory that is experimentally observed!

Have we cornered large $N$ QCD?  On further scrutiny, the spectrum of our solution appears to be too sparse: we find no evidence for the daughter Regge trajectories that would be expected in QCD.  Some caution is in order here because spectrum extraction from the numerical data is quite subtle -- in particular the naive output from the semidefinite solver needs to be interpreted with great care. We discuss several logical possibilities in section \ref{sec:results}. The most optimistic scenario is that by dramatically increasing the number of constraints one would eventually see that the extremal solution contains daughter trajectories. Alternatively, we may have stumbled upon a curious solution that either consists of a single curved trajectory 
(possibly with additional states at very high scale)
or where daughter trajectories kick in at spin $J \gtrsim 10$. It is perhaps not surprising that maximizing the normalized spin-two on-shell coupling may lead to a solution with as sparse a spectrum as possible. 
 In QCD, the contribution of daughter  trajectories to pion scattering appears to be extremely suppressed, which may explain why we find such good numerical estimates for the meson masses and the $g_{\pi\pi f_2}/g_{\pi\pi \rho}$ ratio.  What is clear is that a very economical set of physical assumptions 
has got us either to the final target, or tantalizingly close.

\bigskip
\noindent
The rest of the paper is organized as follows. In section~\ref{sec:Setup}, we review the construction of positivity bounds for large $N$ pion scattering, developed in~\cite{Albert:2022oes}, with special emphasis on the bounds for on-shell couplings. In section~\ref{sec:theorems}, we derive a series of no-go theorems constraining higher-spin resonances by carefully examining the space of null constraints. Section~\ref{sec:results} contains the bulk of our results. By forcing a spin-two state, we find a new stable kink, which we subsequently compare to experimental results. We then study the extremal spectrum at said kink and juxtapose it with the spectrum of real-world mesons. For completeness, we then locate this novel solution in the space of couplings carved out in~\cite{Albert:2022oes}. We conclude in section~\ref{sec:conclusions} with a brief discussion and future directions.
In appendix~\ref{app:RealWorld}, we extract the three-point on-shell couplings of the rho and $f_2$ mesons to two pions from real-world data. In appendix~\ref{app:alongbound}, we discuss the extremal spectrum along the exclusion boundary
where the $f_2$ coupling is maximized. Appendix~\ref{app:LS} contains a discussion of some variations of the Lovelace--Shapiro amplitude which make it compatible with our assumptions.

\section{Setup}\label{sec:Setup}
To make this paper self-contained and fix notations,
we review in this section the basic setup of~\cite{Albert:2022oes}.

\subsection{Generalities of pion scattering}
We consider $2\to 2$ scattering of massless pions at large $N$. The structure of the corresponding amplitude is well known and was extensively reviewed in \cite{Albert:2022oes}. Here we briefly review the setup,  to establish notations and make the paper relatively self-contained. At leading non-trivial large $N$ order, only diagrams with disk  topology contribute. This implies that the amplitude admits the following standard parametrization
in terms of single-traces of the flavor $\mathfrak u(N_f)$ generators,
\begin{align} \label{eq:Tabcd}
{\mathcal T}_{abcd}  = \,& 4\left[\tr{T_aT_bT_cT_d} + \tr{T_aT_dT_cT_b} \right] M(s, t) \nonumber \\ 
+\,& 4\left[\tr{T_aT_cT_dT_b} + \tr{T_aT_bT_dT_c} \right] M(s, u)\nonumber \\
+\,& 4\left[\tr{T_aT_dT_bT_c} + \tr{T_aT_cT_bT_d} \right] M(t, u)\,.
\end{align}
The flavor-ordered amplitude $M(s,u)$ is a function of the Mandelstam invariants alone, which (in ``all-incoming'' conventions) we define by
\be
s = - (p_1 + p_2)^2 \, , \quad t = - (p_2 + p_3)^2\, , \quad u = - (p_1 + p_3)^2 \,.
\ee
Given the structure of traces in \eqref{eq:Tabcd}, invariance of $\mathcal T_{abcd}$ under the exchange of any of the external pions implies that $M(s,u)$ is $s\leftrightarrow u$ crossing symmetric (but not fully $s\leftrightarrow t \leftrightarrow u$ symmetric). The analytic structure of $M(s,u)$ is under good control at large $N$. It is a meromorphic function of $s$ and $u$ with poles in the physical $s$- and $u$-channels. The would-be $t$-channel poles come from exchanges that are suppressed at large $N$, as they do not arise in the disk topology -- this is the so-called Zweig's or OZI rule~\cite{Okubo:1963fa,Zweig:1964jf,Iizuka:1966fk}.

The assumption of unitarity for the full amplitude ${\mathcal T}_{abcd}$ implies that the imaginary part of  $M(s,u)$ admits a partial wave expansion
\begin{equation} \label{eq:ImM}
\text{Im} \, M(s,u) = \sum_{J} n_J \rho_J(s)\, \legP_J\left(1+\frac{2u}{s}\right) \,,
\end{equation}
with positive spectral density $\rho_J(s)\geq 0$ in the physical region. Unitarity usually also implies an upper bound on the spectral density, but there is no meaning to it for large-$N$ scattering amplitudes -- all meson interactions are suppressed by inverse powers of $N$, so $\rho_J(s)\sim \frac{1}{N}$. While such a decomposition holds in any dimension, here we will restrict to four spacetime dimensions for the purposes of comparing with real-world results. In $4d$, the polynomials $\legP_J\left(x\right)$ are Legendre polynomials and the normalization is conventionally chosen as $n_J = 16\pi (2J+1)$ \cite{Correia:2020xtr,Buric:2023ykg}.

At low energies, pion scattering admits an effective field theory description in terms of the familiar chiral Lagrangian
\begin{align} \label{eq:Lch}
    \mathcal L_\text{Ch} =&\, -\frac{f_\pi^2}{4}\tr{\partial_\mu U \partial^\mu U^\dagger}\\
	&+ \kappa_1\tr{(\partial_\mu U  \partial^\mu U^\dagger)^2}
	+ \kappa_2\tr{\partial_\mu U  \partial_\nu U^\dagger  \partial^\mu U\partial^\nu U^\dagger} +  \ldots \,,\nonumber
\end{align}
where $U(x) = \exp\left[i \frac{2}{f_\pi} T_a \pi^a(x) \right]$, and $\kappa_i$ are unfixed Wilson coefficients. This effective theory becomes weakly coupled as $N\to \infty$ as all interaction vertices scale with inverse powers of $N$. At the level of the amplitude, the EFT is simply manifested as a Taylor expansion at low momenta,
\begin{equation}\label{eq:Mlow}
    M(s, u)  \approx g_{1, 0} (s+u) +g_{2, 0} (s^2 + u^2) +2g_{2, 1} \, su + \dots.
\end{equation}
where the low-energy coefficients $g_i$ are in one-to-one correspondence with the Wilson coefficients in \eqref{eq:Lch}. In particular,
\begin{equation}\label{eq:matching}
    g_{1,0} = \frac{1}{2f_\pi^2}\,, \qquad g_{2,0} = \frac{ 2\kappa_1 + 4\kappa_2 }{2 f_\pi^4 }\,, \qquad g_{2,1} = \frac{ 4\kappa_2 }{ f_\pi^4 }\,.
\end{equation}
The radius of convergence of this expansion is given by the location of the first pole in $M(s,u)$, which we denote by $s=M^2$. This defines the cutoff at which the EFT \eqref{eq:Lch} breaks down. For large $N$ QCD, the first exchanged meson in pion scattering is the rho, and so $M^2=m_\rho^2$.

\subsection{Positivity bounds from dispersion relations}
The high-energy behavior of QCD-like amplitudes in the Regge limit ($|s|\to \infty$, fixed $u\lesssim 0$) is controlled by the intercept $\alpha_0(0)$ of the leading Regge trajectory. This is the trajectory of the rho, and since it is a \textit{massive} spin-one particle, $\alpha_0(0)<1$. This allows us to write down dispersion relations with at least one subtraction. There are two independent sets of dispersion relations, dubbed SU and ST in \cite{Albert:2022oes}:
\begin{equation}\label{eq:disprel}
    \text{SU:}\quad \frac{1}{2\pi i } \oint_\infty ds'\, \frac{M(s', u)}{s'^{k+1}}=0\,, \qquad 
    \text{ST:}\quad \frac{1}{2\pi i } \oint_\infty ds'\, \frac{M(s', -s'-u)}{s'^{k+1}}=0\,,
\end{equation}
where $k=1,2,\ldots$. Shrinking the contour then links the pole at the origin, where we can use the EFT expansion \eqref{eq:Mlow} and a cut above the cutoff $M^2$, where we plug the partial wave expansion \eqref{eq:ImM}.

Following the by-now-standard methods of \cite{Caron-Huot:2020cmc}, one can then derive sum rules expressing the low-energy coefficients from \eqref{eq:Mlow} as averages over high-energy data. For the first three coefficients, one finds
\begin{equation}\label{eq:sumrules}
    g_{1,0} = \avg{\frac{1}{m^2}}\,,\qquad
    g_{2,0} = \avg{\frac{1}{m^4}}\,,\qquad
    g_{2,1} = \avg{\frac{J(J+1)}{2m^4}}\,,
\end{equation}
where the high-energy average is defined by
\begin{equation} \label{eq:HEavg}
 \avg{(\cdots)} \equiv
 \frac{1}{\pi} \sum_{J} n_J
 \int_{M^2}^\infty  \frac{dm^2}{m^2}  \rho_J(m^2)\;(\cdots)\,.
\end{equation}
Exploiting crossing symmetry, one further finds two infinite sets of null constraints $\mathcal{X}_{n , \ell},\mathcal{Y}_{n , \ell}$, whose high-energy averages vanish exactly \cite{Caron-Huot:2020cmc, Tolley:2020gtv}. The general expressions can be found in \cite{Albert:2022oes}, here we only quote (in arbitrary normalization) the first ones for later reference
\begin{align}\label{eq:NC}
	m^4 \mathcal{Y}_{2,1}&= 2\left((-1)^J-1\right)+\mathcal{J}^2\,,\\
	m^6 \mathcal{Y}_{3,1}&= 6\left((-1)^J-1\right)+2\left(1-2(-1)^J\right)\mathcal{J}^2\,,\nonumber \\
	m^6 \mathcal{X}_{3,1}&=-6\mathcal{J}^2+\mathcal{J}^4\,, \nonumber \\
        &\cdots \nonumber
\end{align}
where $\mathcal{J}^2\equiv J(J+1)$ is the $SO(3)$ Casimir. It will be an important fact for the interpretation of our results that exchanged states with $J=0$ do not enter the null constraints -- this is an immediate consequence of the need to make at least one subtraction to write valid dispersion relations. On the other hand, scalar states  contribute to the sum rules for the $g_{n, 0}$ low-energy couplings, notably to the one for the lowest coupling $g_{1, 0}$, see (\ref{eq:sumrules}). 

These data allow one to write down a semidefinite problem which can then be solved with a software like \texttt{SDPB}~\cite{sdpb} to derive optimized bounds for normalized ratios of EFT couplings. In particular, \cite{Albert:2022oes} carved out the allowed region in the space of couplings
\begin{equation}\label{eq:gtilde}
    \tilde g_2 \equiv \frac{g_{2,0} M^2}{g_{1,0}}\,, \qquad \tilde g_2' \equiv \frac{2g_{2,1} M^2}{ g_{1,0}}\,.
\end{equation}
At large $N$, we can only bound ratios of EFT couplings because they all scale as $g_{n,\ell}\sim \frac{1}{N}$. (This is precisely what makes the EFT weakly coupled and allows us to neglect EFT loops.) The ratio is then made dimensionless by suitable powers of the cutoff $M^2$. The focus of this paper will be on on-shell three-meson couplings, rather than the four-pion effective couplings of \eqref{eq:Mlow} (coming from integrating out the heavy mesons in the spectrum). To access these couplings, however, we first need to refine our low-energy effective theory.

\subsection{Refining the low-energy EFT}\label{sec:rfinedEFT}

We can push the cutoff $M^2$ higher to extend the domain of validity of our EFT by integrating in new resonances. If we integrate in the first $n$ resonances, the new EFT becomes valid for energies up to the mass of the $n+1$ resonance, which we denote by ${\widetilde M}^2$.
At the level of the amplitude, this is done by including the explicit poles of the corresponding exchanges,
\begin{equation}\label{eq:Mlow-exch}
    M(s,u)\approx \sum_{X=1}^n g_{\pi\pi X}^2
    \left(\frac{m_X^{2} \;\legP_{J_X}\left(1+\frac{2u}{m_X^2}\right)}{m_X^2-s}
    +(s\leftrightarrow u)
    \right)
    + \text{analytic}\,.
\end{equation}
Here $X$ runs over the first $n$ exchanged mesons that we choose to integrate in. In \cite{Albert:2022oes} only the first resonance, the rho meson (with spin $J=1$), was included. Here we will also include the next spin-two exchange: the so-called $f_2$ meson.\footnote{
There are standard naming conventions for the mesons, reviewed e.g.\ in 
Appendix A of \cite{Albert:2022oes}.
For $N_f=2$, the mesons with quantum numbers $J^{PC}=J^{++}$ that are $SU(2)$ isospin triplets are called $a_J$, whereas
the $J^{++}$ isospin singlets are called $f_J$. The
selection rules of the strong interactions imply that it is the $f_2$ (rather than $a_2$) to be exchanged in 2 $\to$ 2 pion scattering. Note however that 
at large $N$ this distinction becomes immaterial, because the flavor symmetry gets enhanced to $U(N_f)$ and the different isospin projections combine into one multiplet -- the adjoint of $U(N_f)$.  
}
The ``analytic'' piece in \eqref{eq:Mlow-exch} can be parametrized as a crossing-symmetric expansion similar to \eqref{eq:Mlow}, but with \textit{different coefficients}. The amplitude \eqref{eq:Mlow} is recovered upon Taylor-expanding (a.k.a.\ integrating out) the poles in \eqref{eq:Mlow-exch} around $s,u\sim 0$.

One should view \eqref{eq:Mlow-exch} as the amplitude arising from an extension of the chiral Lagrangian \eqref{eq:Lch} incorporating new fields for the $X$ resonances, which we will not bother writing. In particular, $g_{\pi\pi X}$ captures the three-point interaction between two pions and a meson $X$ (in some suitable normalization). We spell out this interaction in detail in appendix~\ref{app:RealWorld}, where we also extract the couplings $g_{\pi\pi\rho}$ and $g_{\pi\pi f_2}$ from real-world data.

With this more refined EFT, we can now proceed as before and write down the dispersion relations \eqref{eq:disprel} where we now use \eqref{eq:Mlow-exch} below the new cutoff $\widetilde M^2$ and the partial wave expansion for the cuts above it, where we remain agnostic about the spectrum. Now the contour integral will step on the $s=m_X^2$ poles, which will give us access to the residues $g_{\pi\pi X}^2$. In practice, this is straightforward to implement by keeping the new poles in the high-energy side of dispersion relations. We redefine the spectral density to include a delta function for each of the low-lying exchanges, and remain unknown (but positive) above the new cutoff $\widetilde M^2$,
\begin{equation}
    \rho_J(s) = \sum_X g_{\pi\pi X}^2\, \frac{\pi m_X^2}{n_J}\delta(s-m_X^2) \delta_{J,J_X} + \widetilde \rho_J(s)\,.
\end{equation}
Here $\widetilde \rho_J(s)\geq 0$ has support only for $s\geq \widetilde M^2$. Plugging this into the high-energy averages \eqref{eq:HEavg} simply produces new terms due to the explicit exchanges:
\begin{equation}
    \avg{F(m^2,J)}_{m^2\geq M^2} = \sum_X g_{\pi\pi X}^2F(m_X^2,J_X) + \avg{F(m^2,J)}_{m^2\geq \widetilde M^2}\,.
\end{equation}
For example, when including only the $X = \rho, f_2$ exchanges, the sum rules \eqref{eq:sumrules} become
\begin{align}
    g_{1,0} =\,& \frac{g_{\pi\pi\rho}^2}{m_\rho^2} + \frac{g_{\pi\pi f_2}^2}{m_{f_2}^2} + \avg{\frac{1}{m^2}}_{m^2\geq \widetilde M^2}\,,\\
    g_{2,0} =\,& \frac{g_{\pi\pi\rho}^2}{m_\rho^4} + \frac{g_{\pi\pi f_2}^2}{m_{f_2}^4} + \avg{\frac{1}{m^4}}_{m^2\geq \widetilde M^2}\,,\nonumber \\
    g_{2,1} =\,& \frac{g_{\pi\pi\rho}^2}{m_\rho^4} + 3\frac{g_{\pi\pi f_2}^2}{m_{f_2}^4} + \avg{\frac{J(J+1)}{2m^4}}_{m^2\geq \widetilde M^2}\,.\nonumber
\end{align}
We note that the cutoff $\widetilde M^2$ here can be chosen to depend on $J$, if one wishes to integrate in mesons of different masses spin by spin.

With these new sum rules (and null constraints) including the on-shell couplings $g_{\pi\pi X}^2$ explicitly, we may now proceed to write down a semidefinite problem to put bounds on these couplings. This was carried out in \cite{Albert:2022oes} for the rho coupling, and it is entirely analogous to the algorithm to bound OPE coefficients in the CFT bootstrap \cite{Caracciolo:2009bx}. 
We refer the reader to these references for the explicit formulation of the problem.
What we will emphasize here is that we are again only allowed to bound \textit{ratios of couplings} that cancel the $N$ dependence as $N\to \infty$. All three-meson couplings scale as $g_{\pi\pi X} \sim\frac{1}{\sqrt{N}}$ at large $N$, so one might try to look for bounds on $g_{\pi\pi f_2}^2/g_{\pi\pi\rho}^2$. To directly bound such a ratio is difficult because the sum rule for $g_{\pi\pi\rho}^2$ is not manifestly positive-definite.\footnote{See section 7.3 of \cite{Albert:2023jtd} for a discussion of how this presents an obstruction in carrying out the semidefinite program.}
What is straightforward to bound, instead, are the dimensionless ratios\footnote{Compared to \cite{Albert:2022oes}, we are using a different normalization for the $g_{\pi\pi\rho}$ coupling in \eqref{eq:Mlow-exch}. We make up for this fact here so the ratio $\tilde g_\rho^2$ matches the one there.}
\begin{equation}\label{eq:grhogf2}
    \tilde g_\rho^2 \equiv \frac{g_{\pi\pi\rho}^2}{g_{1,0}m_\rho^2}\,, \qquad
    \tilde g_{f_2}^2 \equiv \frac{g_{\pi\pi f_2}^2}{g_{1,0}m_\rho^2}\,.
\end{equation}
The factor of $g_{1,0}$ cancels the $N$ dependence, and we use powers of the mass of the rho (as opposed to the cutoff $\widetilde M^2$) to make the ratio dimensionless. We will come back to these ratios in section \ref{sec:results}.

\section{Higher spins and null constraints}

\label{sec:theorems}

In this section we 
derive a few spectral no-go theorems that constrain the space of solutions of our bootstrap equations. We rely only on null constraints. As we have explained, since valid dispersion relations need at least one subtraction, null constraints are insensitive to $J=0$ states. Our discussion will thus be entirely agnostic to the presence of intermediate scalar states.

\subsection{Geometry of the null constraints: a graphical bootstrap}

Here we give a graphical proof of several important facts, using the following strategy. We select two particular combinations of null constraints $n^{(1)}, n^{(2)}$ (different each time),
 \begin{equation}
    \begin{pmatrix}
        0\\0
    \end{pmatrix} =  \frac{1}{\pi} \sum_{J} n_J
 \int_{M^2}^\infty  \frac{dm^2}{m^2}  \rho_J(m^2) \underbrace{\begin{pmatrix}
        n^{(1)}_J(m^2)\\n^{(2)}_J(m^2)
    \end{pmatrix}}_{\vec v_J(m^2)}\,.
\end{equation}
We then plot the contribution
of a state with mass $m$ and spin $J$ to these null constraints as a (properly normalized) vector in the plane $(n^{(1)},n^{(2)})$. For a given $J$, the vectors $\vec v_J(m^2)$, parametrized by $m^2\geq M^2$, describe a curve. 
As the vectors $\vec v_J(m^2)$ must sum to zero, they cannot be entirely contained in a half plane. Thus, whenever we find a spectrum of states which only produces curves lying on the same half-plane we can claim that that spectrum is inconsistent.

\begin{figure}[h]
\centering
\includegraphics[width=0.7\textwidth]{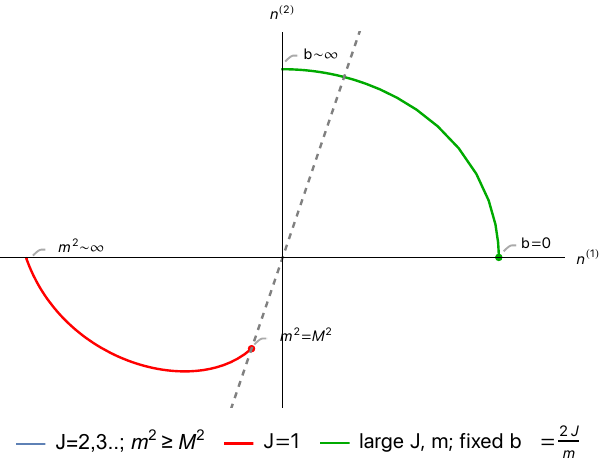}
\caption{The figure shows two particular directions in the space of  null constraints, namely $n^{(1)}=\mathcal{Y}_{3,1}-\mathcal{Y}_{2,1}$ and $n^{(2)}=\mathcal{X}_{3,1}$, for $J=1$ and mass $m^2\geq M^2$ (red line). Vectors have been properly normalized to help visualization. The green line corresponds to the limit of null constraints $m, J\rightarrow\infty$ with fixed $b=2J/m$. Resonances with $J=1$ alone lie on the same half-plane, thus making it impossible for a linear combination with positive coefficients to sum to zero. This becomes possible when including resonances at infinity with suitable values of $b$.
}
    \label{fig:only-spin-1}
\end{figure}

Let us start with a simple realization of this idea. We display in figure~\ref{fig:only-spin-1} a pair of null constraints such that the curve (in red) spanned by the states with $J=1$ is entirely contained in the lower left quadrant. We conclude that spin-one states alone cannot produce an amplitude consistent with our dispersion relations. This is expected since they give linearly growing amplitudes in the Regge limit, which are marginally inconsistent with  (\ref{mesonRegge}). As  shown in~\cite{Albert:2022oes}, such mild violation can be fixed by adding states at arbitrary large mass. This suggests that the large $m$ regime plays a crucial role. The asymptotic behavior of the null constraints at large $m$ also depends on the value of $J$. We can parametrize the asymptotic regime in terms of the impact parameter $b = \frac{2J}m$.
In the large-$m$ limit, $b=0$ represents the behavior of null constraints with fixed spin, while finite values of $b$ correspond to large $J, m$ with fixed ratio. Finally $b\rightarrow\infty$ represents states with spin growing faster than $m$.\footnote{This regime was shown to play a fundamental role in the EFT bootstrap already in \cite{Caron-Huot:2021rmr,Caron-Huot:2022ugt,Henriksson:2021ymi,McPeak:2023wmq}.} As expected, we see in figure~\ref{fig:only-spin-1}  that including states at large $m,J$ with fixed $b$ (green line) allows to satisfy the constraints, since the two curves now span the whole plane.

\begin{figure}[h]
\centering
\includegraphics[width=\textwidth]{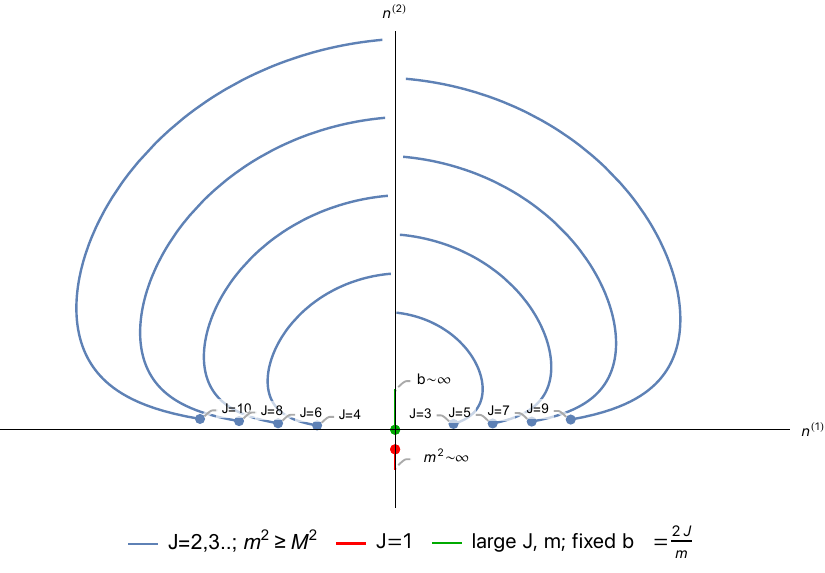}
\caption{The figure shows two particular directions in the space of  null constraints, ${n^{(1)}=\mathcal{Y}_{4,1}}$ and $n^{(2)}=\mathcal{X}_{3,1}$, for various values of the spin $J$ and the mass $m^2\geq M^2$ (red: $J=1$, blue: $J>1$, green: asymptotic). Dots correspond to the minimal value of $m^2$. Again the vectors have been
properly normalized to help visualization. If we forbid the presence of $J=1$ states, all other points lie on the same half-plane.}
    \label{fig:no-spin-1}
\end{figure}
Interestingly, we can show that the presence of spin-one states is actually {\it necessary} for consistency, 
see figure~\ref{fig:no-spin-1}. The combination plotted on the vertical axis receives positive contributions from all states except from $J=1$, which instead produces the  negative contribution needed to balance the sum.

\begin{figure}[h]
\centering
\includegraphics[width=0.8\textwidth]{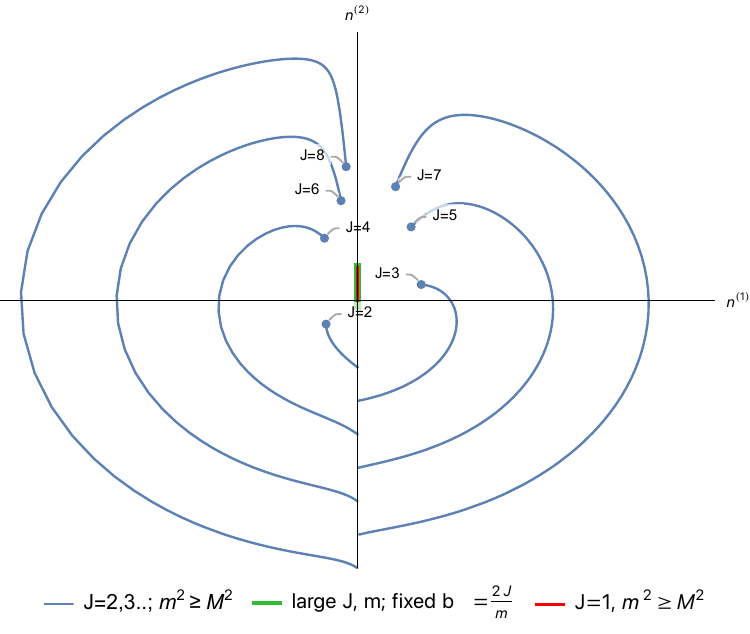}
\caption{The figure shows two particular directions in the space of  null constraints, \mbox{$n^{(1)}=\mathcal{Y}_{3,1}$} and $n^{(2)}=4\mathcal{Y}_{2,1}+\mathcal{X}_{3,1}$, for various values of the spin $J$ and the mass $m^2\geq M^2$ (red: $J=1$, blue: $J>1$, green: asymptotic). Again the vectors have been
properly normalized to help visualization. Dots correspond to the lowest valued $m^2=M^2$. A $J=2$ state cannot be compensated by any $J=1,2$ state nor by resonances at infinity: in order to compensate it one needs at least one state with odd $J>2$.}
    \label{fig:only-spin-1-2}
\end{figure}

One could try to imitate the logic of figure \ref{fig:only-spin-1} for higher-spin states, but
there is a qualitative difference in the way  $J=1$ and $J>1$
 states contribute the null constraints.  Indeed, one can find combinations of null constraints that receive contribution from $J >1$ states of finite mass, but {\it not} from either $J=1$ states or
the asymptotic regime at fixed $b$. For instance, at $n_\text{max}=3$ one has
\begin{align}
    \begin{pmatrix}
        \mathcal{Y}_{2,1}\\\mathcal{Y}_{3,1}\\\mathcal{X}_{3,1}
    \end{pmatrix}\Bigg|_{J=1}= 
     \begin{pmatrix}
        2 m^2\\0\\-8
    \end{pmatrix},\qquad
    \begin{pmatrix}
        \mathcal{Y}_{2,1}\\\mathcal{Y}_{3,1}\\\mathcal{X}_{3,1}
    \end{pmatrix}\Bigg|_{J=2}= 
     \begin{pmatrix}
        -6m^2\\-24\\0
    \end{pmatrix},\qquad
        \begin{pmatrix}
        \mathcal{Y}_{2,1}\\\mathcal{Y}_{3,1}\\\mathcal{X}_{3,1}
    \end{pmatrix}\Bigg|_{m,J\rightarrow\infty}\propto
     \begin{pmatrix}
        -1\\0\\ \frac{b^2}4
    \end{pmatrix}.
\end{align}
(In fact all spins with $J>1$ contribute to $\mathcal{Y}_{3,1}$).  This means that a spectrum containing any number of states with $J>1$ (for given $J$) cannot be made consistent  by only adding $J=1$ states or states at infinite mass. This is shown in figure~\ref{fig:only-spin-1-2}. We can also observe that even or odd spins alone are inconsistent, as they all lie on the same half plane. The argument can be iterated: considering for instance the plane $(n^{(1)},n^{(2)})=(\mathcal{Y}_{5,2},\mathcal{X}_{5,2})$ one can show that the economic choice of only adding states with $J=3$ would still be inconsistent.

Finally, we can take a similar approach to test if particular choices of spectra are consistent. For example, consider the spectrum corresponding to a single linear Regge trajectory, with a single state per spin $J$ and mass given by the relation
\begin{equation}
    m^2_\text{Regge}(J)=(J-2)(m^2_{f_2}-m_\rho^2)+m_{f_2}^2 \,.
\end{equation}
 (The $J=1$ and $J=2$ states have been given suggestive names).
For instance one could fix $m_{f_2}^2 = 3m_\rho^2$, which is the slope of the leading Regge trajectory in the Lovelace--Shapiro amplitude \eqref{eq:LSamplitude}. By choosing carefully the combinations of null constraints one can show that a single linear Regge trajectory is inconsistent. One would need to at least include states at infinity {\it outside} the trajectory (finite $b$), since on the trajectory \mbox{$b\sim 2J/m_\text{Regge}\rightarrow \infty$}.

To summarize, a simple graphical bootstrap that leverages clever choices of null constraints allowed us to show that:
\begin{enumerate}
    \item[(i)] $J=1$ states must necessarily be present;
    \item[(ii)]  $J=1$ states alone are not consistent but can be compensated by adding states at infinite mass;
    \item[(iii)] Including any additional state with 
    an even (odd) spin
    $J>1$ requires more states with odd (even) spin and \emph{finite} mass;
    \item[(iv)] A single linear Regge trajectory is not consistent, but it could in principle be made consistent by including states with finite impact parameter $b$.
    \end{enumerate}

\section{A novel extremal solution}\label{sec:results}

In the program initiated in \cite{Albert:2022oes} and continued in \cite{Albert:2023jtd,Fernandez:2022kzi,Ma:2023vgc,Li:2023qzs}, the ultimate goal is to corner large $N$ QCD. This means finding a solution to the bootstrap  with an infinite number of states arranging in Regge trajectories, which match the physical meson spectrum. To date, no solution
that looks even qualitatively like large $N$ QCD
has made an appearance in the dual methods employed in these explorations.\footnote{Regge trajectories have  showed up in the non-perturbative S-matrix bootstrap \cite{Acanfora:2023axz,Guerrieri:2023qbg}.} In fact, most of the solutions saturating positivity bounds are far from even being \textit{physical}, involving -- for example -- an infinite tower of spins at a given mass.

The upshot is that our assumptions are too weak, 
allowing for artificial solutions to the bootstrap constraints.  In \cite{Albert:2022oes}, a first step to 
inject further physical input was taken, by insisting that the pion amplitude should include the exchange of
the lightest massive meson  (the spin-one rho meson). While this provided some new insights, it still did not bring about QCD-like solutions. We have just recalled why: a single spin-one exchange can be UV-completed by contributions at infinity (figure \ref{fig:only-spin-1}). This solution was already identified in \cite{Albert:2022oes}, and it involved an infinite tower of higher spins at the same mass.

The discussion in the previous section suggests how to go beyond this paradigm. We need to enforce the existence of an intermediate  state with spin $J >1$. 
In large $N$ QCD, 
the first higher-spin meson exchanged in pion scattering is the $f_2$, a massive spin-two state.
We have seen that as soon as a state with $J=2$ is present, any bootstrap solution will necessarily involve a higher odd-spin state at a higher (but \textit{finite}) mass. By iterating the same argument, one can show that an infinite Regge trajectory is needed. This is very natural, we are assuming a better-than-\textit{spin-one Regge behavior}, granted by the intercept of the leading Regge trajectory $\alpha_0(0)<1$. A spin-one exchange fails to satisfy this behavior, but only marginally, which can be compensated by infinitely-heavy states. For a spin-two (or higher) exchange, in contrast, the Regge behavior is much worse, and one needs full Regge trajectories to make up for it.

Before we proceed, let us comment on a subtle but important point.
At large $N$, all mesons arrange in degenerate multiplets of $U(N_f)$, concretely in the adjoint representation. So, in principle, we need not distinguish the $f_2$ (an isospin singlet) from the $a_2$ (its isospin triplet counterpart). At finite $N$, these multiplets break into different representations of $SU(N_f)$ (singlet and adjoint), which are further broken in the real world by the flavor symmetry being only \textit{approximate}. This breaks the degeneracy between the $f_2$ and $a_2$ mesons. In the scattering of the full $U(N_f)$ pion multiplet (which in particular includes the $\eta'$ meson), both of them occur. But if we restrict to the scattering of $SU(2)$ pions, only the $f_2$ couples.\footnote{See the discussion in appendix A of \cite{Albert:2022oes} for a discussion on selection rules for pion scattering.} When comparing to real-world data, we will do such a restriction, and we will therefore focus only on data for the $f_2$ (rather than the $a_2$) meson.

\subsection{Forcing a spin-two state: the $f_2$ kink}

\begin{figure}[h]
\centering
\includegraphics[width=1.02\textwidth]{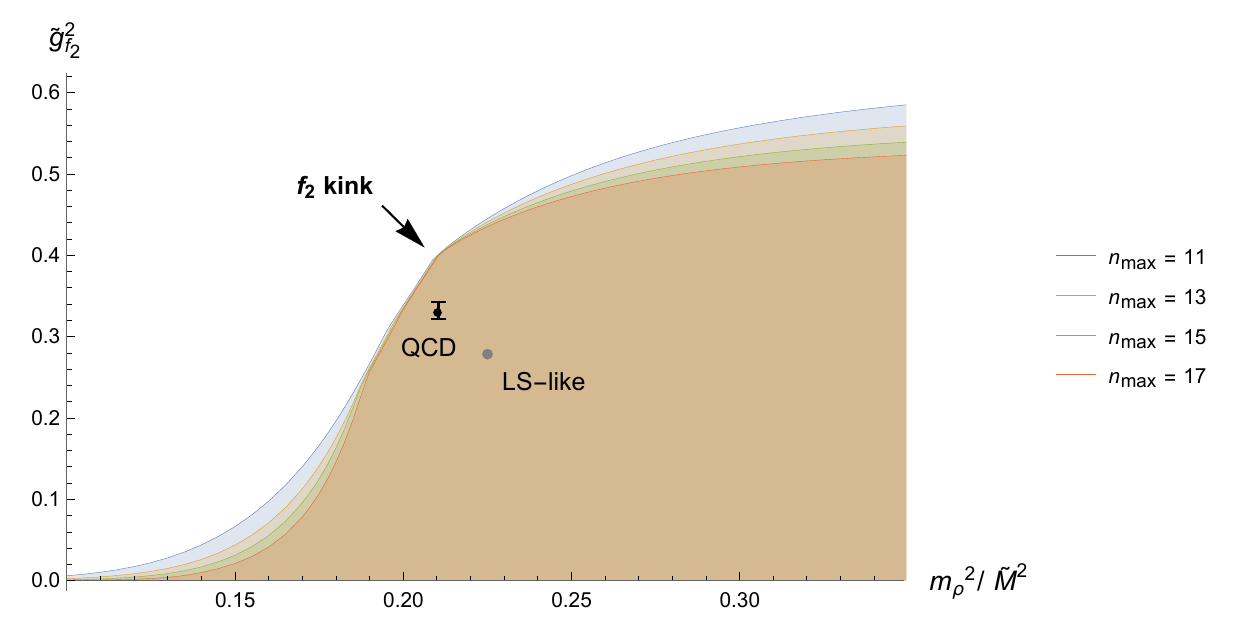}
\caption{Upper bound on the $f_2$ on-shell coupling as a function of the cutoff. The different colors indicate the ``Mandelstam order'' of the highest included null constraint. The bounds have not yet converged except for a stable kink around $m_\rho^2/\widetilde M^2\approx 0.21$ (and a neighboring region).
The black dot with uncertainty in the coupling direction corresponds to real-world QCD \eqref{eq:physicalCouplings}, and the gray dot corresponds to the linear-trajectory amplitude with scalars removed \eqref{eq:Linear-minus-scalars}.
}
    \label{fig:kink-plot}
\end{figure}

So the task is clear. We should include both the rho and the $f_2$ mesons as explicit poles in the amplitude. There are two free spectral parameters: the ratio of the two meson masses, which we fix to the value for real-world QCD,
\begin{equation}\label{eq:PhysicalValue}
    \frac{m_{f_2}^2}{m_\rho^2} = (1.65)^2\,,
\end{equation}
and the ratio to the cutoff $m_\rho^2/\widetilde M^2$, which we will scan over.\footnote{
In real-world QCD there are other spin-one and spin-zero mesons kicking in \textit{before} the $f_2$ (lying on subleading Regge trajectories), and there might be further states below the cutoff $\widetilde M^2$. It is possible to derive bounds with assumptions accommodating these possibilities, for instance by allowing spin $J=0,1$ states to lie anywhere above the rho mass $m_\rho$, spin $J=2$ states above the $f_2$ mass, and consequently only using $\widetilde M$ for $J\geq3$. The result for extremizing the $f_2$ coupling does not change under such milder assumptions; on the other hand, as seen in figure~\ref{fig:EFTcouplings}, the bound for the EFT couplings do.
}
As discussed in section~\ref{sec:rfinedEFT}, with this setup we can bound \textit{on-shell couplings}, which are more interesting than the usual low-energy couplings of \eqref{eq:Lch}. Figure \ref{fig:kink-plot} shows the upper bound on the normalized $f_2$ coupling $\tilde g^2_{f_2}$ (defined in \eqref{eq:grhogf2}) as a function of the cutoff.

The first thing that we notice in figure \ref{fig:kink-plot} is that the bound tends to zero as $m_\rho^2/\widetilde M^2\to 0$. Reading the plot horizontally, this means that fixing a non-zero $f_2$ coupling, so as to enforce the presence of a spin-two exchange, imposes a (finite) upper bound on the cutoff $\widetilde M^2$ (in units of the mass of the rho). This confirms our expectations -- the heavy masses cannot come in too high if they are to UV-complete a spin-two exchange. Note that this is in stark contrast with the bounds on the rho coupling presented in \cite{Albert:2022oes} (see e.g.\ figure 16 therein). The bound in that case plateaus as $\widetilde M^2\to \infty$, in agreement with the UV completion of the rho meson using only infinitely-heavy masses.

\begin{figure}[h]
\centering
\includegraphics[width=0.95\textwidth]{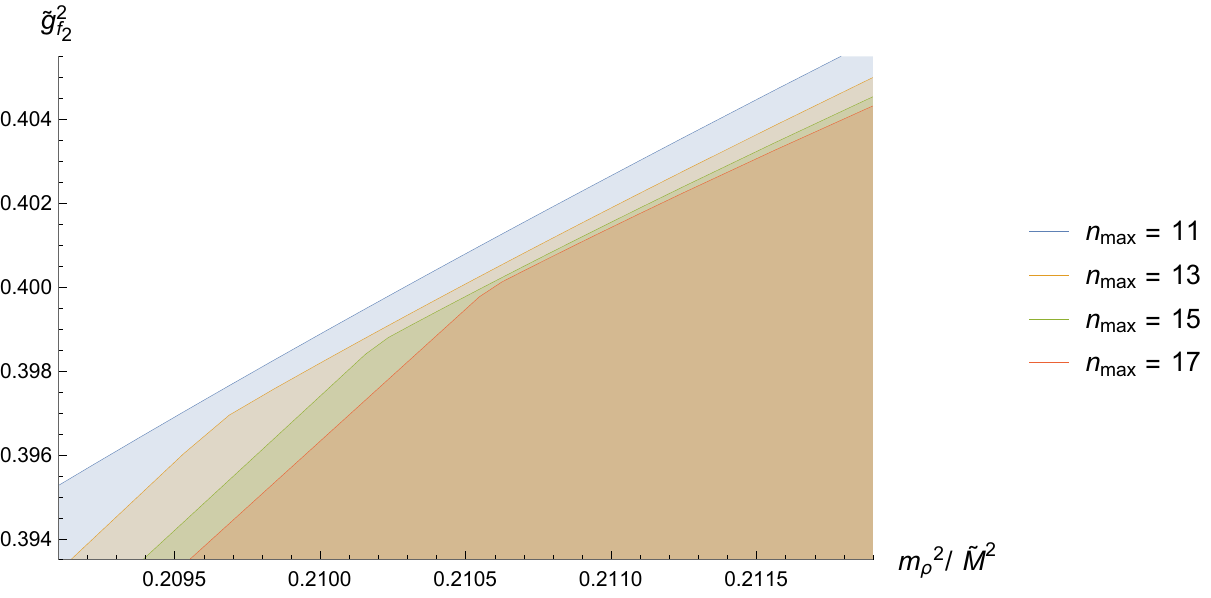}
\caption{Same as figure~\ref{fig:kink-plot}, zooming in on the kink.
}
    \label{fig:kink-plot-zoom}
\end{figure}

The surprising feature of figure \ref{fig:kink-plot} is the appearance of a stable kink. To the right and left of the plot, the bound is still far from having converged in null constraints. Each of the bounds is rigorous but not optimal. Near the center, however, there is a point where the converge is much faster, indicating the proximity to a true solution to the bootstrap. We present a close-up of this region in figure \ref{fig:kink-plot-zoom}, from which we may read the position of the kink,
\begin{equation}
    \frac{m_\rho^2}{\widetilde M^2}\approx 0.2106\,.
\end{equation}
What is the extremal solution at
this kink? The first resource is to turn to the spectrum found by \texttt{SDPB}~\cite{sdpb} but, as we will see below,  to analyze the extremal spectrum requires some care. Instead, we choose to first locate known solutions on this plot, and we defer a discussion on the spectrum until section~\ref{sec:spectrum}.

We note here that we have also checked that picking different values for the ratio
$m_{f_2}/m_\rho$ does not qualitatively change the story; the position of the kink moves smoothly and the extremal solution has the same  general features.  We will perform our detailed analysis with the physical value~(\ref{eq:PhysicalValue}).

\subsection{Comparing to experiment}\label{sec:Experiment}
Let us now compare with experimental results from real-world QCD. The first $J\geq 3$ resonance in QCD is the $\rho_3$, a spin-three meson heavier than the $f_2$. So, for QCD, the cutoff $\widetilde M$ should be identified with $m_{\rho_3}$. The corresponding value is
\begin{equation}
    \frac{m_{\rho_3}^2}{m_\rho^2} \approx 4.747\,,
\end{equation}
which is actually very close to the kink, which has $\widetilde M^2/m_\rho^2\approx 4.748$. Note that this is significantly away from the linear trajectory going through the rho and $f_2$, where the string-like amplitudes lie.\footnote{Note that in real-world QCD, the $(a,\rho)$ and the $(f,\omega)$ trajectories are not completely degenerate.} See appendix \ref{app:LS} for a discussion on a variation of the Lovelace-Shapiro amplitude that is consistent with our assumptions. Its location is marked by a gray dot in figure \ref{fig:kink-plot}. 

The physical on-shell couplings for the meson exchanges in pion scattering can be determined from their decay rates into two pions. This is worked out in detail in appendix \ref{app:RealWorld} for the rho and $f_2$ mesons. Here we simply quote the results,
\begin{equation}
\label{eq:physicalCouplings}
    \tilde g_\rho^2 \simeq 0.504\pm 0.009\,, \qquad \tilde g_{f_2}^2 \simeq 0.329^{+0.013}_{-0.007}\,.
\end{equation}
With these results, we can place real-world QCD in our exclusion plot. It is marked with a black dot (with uncertainty) in figure \ref{fig:kink-plot}. In contrast to the mass, which is quite close to that of the kink, the coupling is well below it. But this is not surprising. For one thing, recall that our setup is blind to scalars, so we are always allowed to subtract them from any given solution. Subtracting them from QCD would keep the on-shell coupling $g_{\pi\pi f_2}^2$ fixed, but decrease $g_{1,0}$, pushing the normalized coupling $\tilde g_{f_2}^2\sim g_{\pi\pi f_2}^2/g_{1,0}$ higher.

\begin{figure}[h]
\centering
\includegraphics[width=0.85\textwidth]{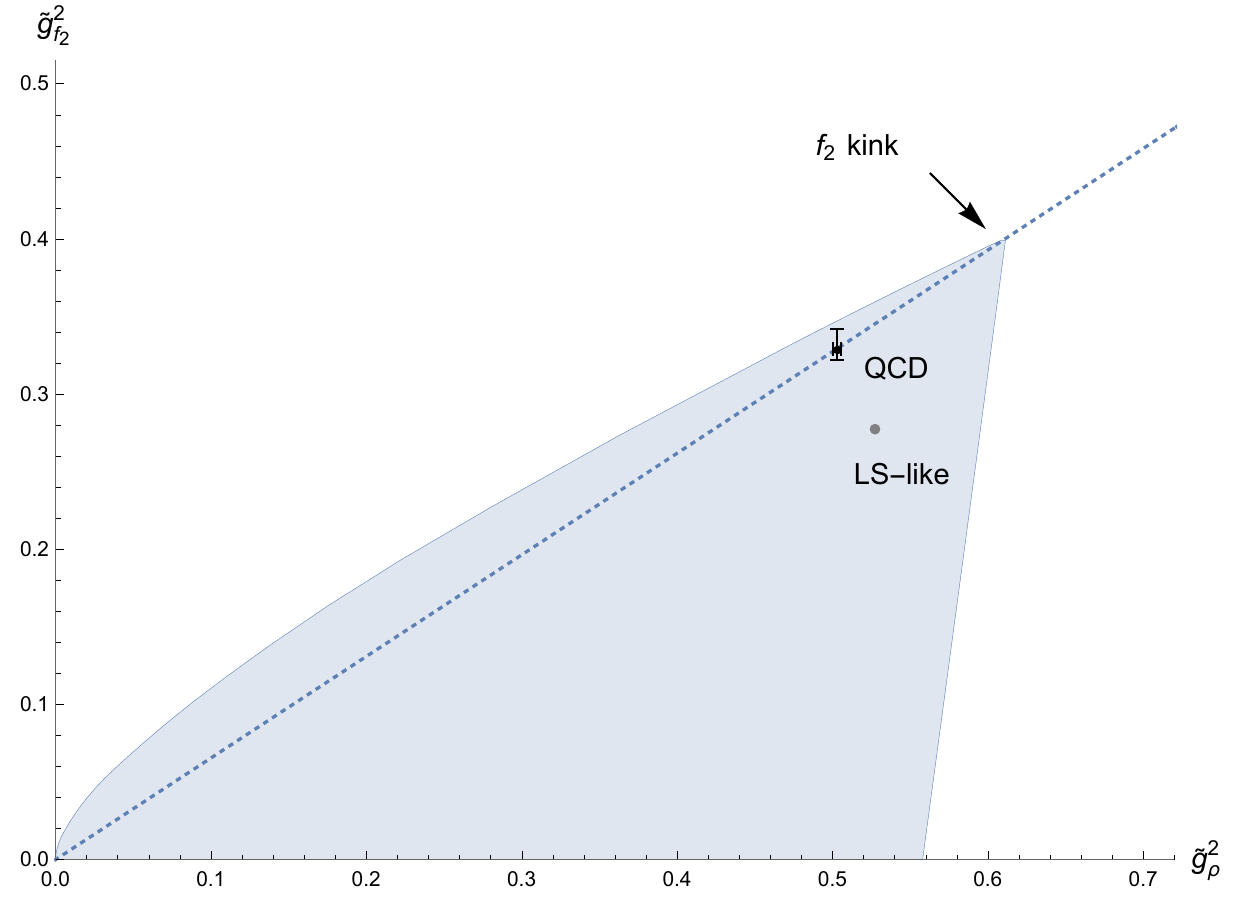}
\caption{Allowed value in the space of the first two on-shell couplings. The black rectangle denotes the values of real-world QCD \eqref{eq:physicalCouplings}. The gray dot denotes the linear amplitude with scalars removed, \eqref{eq:Linear-minus-scalars}. The dotted line is chosen to go though the corner of the allowed region, which coincides with values at the kink. The plot was made at $n_{\mathrm{max}}=15$. 
}
    \label{fig:couplings}
\end{figure}

In light of this, one may hope that the kink corresponds to large $N$ QCD with scalars (and perhaps more) subtracted. A more meaningful observable is then the ratio
\begin{equation}
\frac{\tilde g_{f_2}^2}{\tilde g_\rho^2} = \frac{g_{\pi\pi f_2}^2}{g_{\pi\pi \rho}^2}\,,
\end{equation}
which cancels out the $g_{1,0}$ dependence, susceptible to subtractions. Fixing the cutoff $\widetilde M$ to the horizontal position of the kink, and scanning over $\tilde g_{f_2}^2$, allows one to carve out the allowed region in the space of normalized couplings $\tilde g_\rho^2,\tilde g_{f_2}^2$. This is shown in figure \ref{fig:couplings}. Here, the stable kink of figure \ref{fig:kink-plot} corresponds to the top-right corner, and real-world QCD is again marked on this plot with a black dot (with uncertainty).

While real-world QCD is not too close to the kink, it lies (within uncertainty) just on top of the dashed line representing the ratio $\tilde g_{f_2}^2/\tilde g_\rho^2$ for the kink. The ratio of the rho and $f_2$ on-shell couplings at the $f_2$ kink is thus compatible with experimental QCD! This supports the idea that the $f_2$ kink might correspond to large $N$ QCD but with a sparser spectrum  (such as subtracting scalars), which would decrease $g_{1,0}$ pushing the normalized couplings out all the way to the top-right corner. We will discuss this possibility further when we investigate the spectrum.

\subsection{The spectrum at the $f_2$ kink}\label{sec:spectrum}
We turn to analyze the extremal spectrum at the $f_2$ kink. As always, the spectrum saturating any positivity bound can be extracted from the extremal functional obtained from the semidefinite solver that is being used (\texttt{SDPB}~\cite{sdpb} in our case).\footnote{This is entirely analogous to the extraction of the extremal spectrum in the conformal bootstrap. On-shell couplings can also be retrieved from an \texttt{SDPB} primal-dual optimal solution, just like OPE coefficients in the conformal bootstrap \cite{Komargodski:2016auf,Simmons-Duffin:2016wlq}.
}
Inspecting the spectrum while moving along the boundary of figure \ref{fig:kink-plot}, from left to right, one observes an initial chaotic collection of states localized a the cut off $\widetilde M^2$, which starts organizing into a trajectory that becomes more and more pronounced as one approaches the $f_2$ kink. Past the kink the trajectory flattens out. The $f_2$ kink then appears to be linked to the formation of a Regge trajectory.

\begin{figure}
	\centering
 \begin{subfigure}[b]{0.47\textwidth}
		\includegraphics[width=\textwidth]{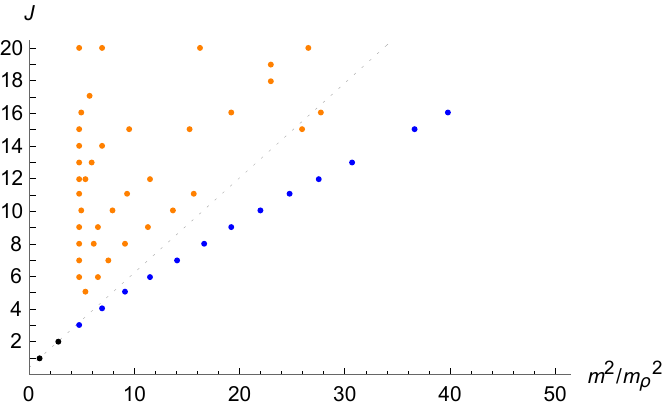}
  \caption{}
  		\label{fig:spectrum-at-kink-A}
  \end{subfigure}
\begin{subfigure}[b]{0.47\textwidth}
		\includegraphics[width=\textwidth]{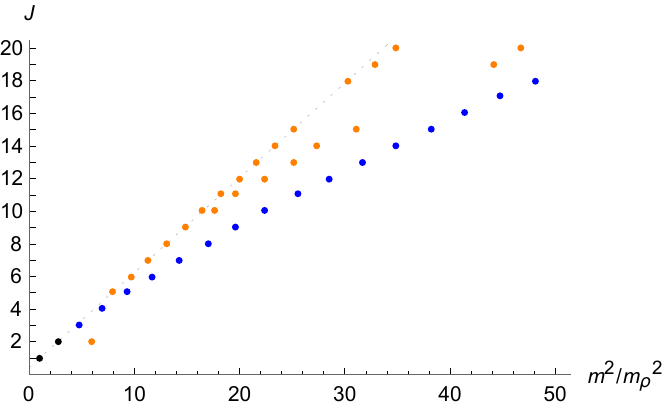}
  \caption{}
  		\label{fig:spectrum-at-kink-B}
   \end{subfigure}
		\caption{The extremal spectra at the point closest to the $f_2$ kink, at $n_{\mathrm{max}}=17$. Left: original assumptions, right: assuming states to the right of the linear extrapolation only for $J\neq3$, and above the $M'$ at $J=3$. As table~\ref{tab:dominant-trajectory} shows, the points above the dominant trajectory have very small relative couplings, and we have displayed them in a different color. Note that any daughter trajectories should lie below the leading trajectory. To  guide the eye, we show the linear extrapolation trajectory passing through $\rho$ and $f_2$ in dashed.}
  \label{fig:spectrum-at-kink}
\end{figure}

Concretely, the spectrum at the $f_2$ kink of figure \ref{fig:kink-plot}, obtained from \texttt{SDPB}, is shown in figure \ref{fig:spectrum-at-kink-A}. We first note that, as we expected, it features a long Regge trajectory (marked in blue) continuing beyond the locations of the rho and $f_2$ mesons, which we introduced by hand. Interestingly, the trajectory bends to the right of the line traced by the first two mesons, which is marked by a gray dashed line in figure \ref{fig:spectrum-at-kink-A}.
Above this main trajectory, the spectrum contains a large collection of states (marked in orange) clustering towards the cutoff $\widetilde M^2$. Similar-looking spectra were reported in previous bootstrap studies \cite{Caron-Huot:2021rmr,Albert:2022oes}. In QCD-like theories, we do expect an infinity of Regge trajectories, but the orange states are in the wrong side of the plot to be interpreted as subleading trajectories. It turns out that all these additional states are spurious; they are  artifacts of the truncations introduced when using numerical solvers like \texttt{SDPB}.

The first indication that these states are spurious is that, compared to the blue states, they are quite unstable under changing the number of null constraints being used. More evidence comes from examining their on-shell couplings. Table \ref{tab:dominant-trajectory} lists the masses and normalized on-shell couplings $\tilde g_X^2\equiv g^2_{\pi\pi X}/(g_{1,0}m_\rho^2)$ for the low-lying states of figure \ref{fig:spectrum-at-kink-A}. The entries in the first column correspond to the (blue) states on the main Regge trajectory. To the right are the remaining (orange) states for every spin. We see that the couplings of the states which fall outside of the main trajectory are suppressed by at least three orders of magnitude compared to the states on it. This shows that the main (curved) Regge trajectory dominates over the remaining states.

\begin{table}
    \caption{Table of the states at the extremal spectrum extracted at $m_\rho^2/\widetilde M^2=0.210625$. Masses are expressed in units where $m_\rho=1$.
}\label{tab:dominant-trajectory}
\resizebox{\textwidth}{!}{
{\small 
\begin{tabular}{|l|ll|llllll|}
\hline
 & \multicolumn{2}{c|}{Dominant state} & \multicolumn{6}{c|}{Other states}
\\\hline
$J$& $m^2$ & $\tilde g_X^2$& $m^2$ & $\tilde g_X^2$& $m^2$ & $\tilde g_X^2$& $m^2$ & $\tilde g_X^2$
\\\hline
3&
4.747774 & 0.33527 & &&&&& \\
4&
 6.902792 & 0.28933 &  &&&&& \\
 5&
 9.181336 & 0.25334 & 5.278811 & 1.515$\times 10^{-4}$ &  &  &  &  \\
 6&
 11.54579 & 0.22174 & 4.747774 & 3.297$\times 10^{-6}$ & 6.582414 & 1.773$\times 10^{-4}$ &  &  \\
 7&
 14.01378 & 0.19857 & 4.835758 & 7.862$\times 10^{-7}$ & 7.581251 & 1.237$\times 10^{-4}$ &  &  \\
 8&
 16.67318 & 0.18599 & 4.747774 & 8.771$\times 10^{-8}$ & 6.235041 & 1.265$\times 10^{-6}$ &
   9.207674 & 1.352$\times 10^{-4}$ \\
   9&
 19.28674 & 0.16358 & 4.808180 & 1.895$\times 10^{-8}$ & 6.571938 & 4.367$\times 10^{-7}$ &
   11.31167 & 2.537$\times 10^{-4}$ \\
   10&
 21.93016 & 0.14912 & 5.019793 & 7.308$\times 10^{-9}$ & 7.879411 & 3.754$\times 10^{-7}$ &
   13.61458 & 4.242$\times 10^{-4}$ \\
   11&
 24.82063 & 0.11649 & 4.825621 & 6.643$\times 10^{-10}\!\!$ & 9.289181 & 1.875$\times 10^{-6}$ &
   15.69828 & 1.554$\times 10^{-4}$ \\
   12&
 27.53345 & 0.10811 & 4.747774 & 8.380$\times 10^{-11}\!\!$ & 5.390215 & 7.235$\times 10^{-11}\!\!$ &
   11.48907 & 7.067$\times 10^{-6}$
   \\\hline
\end{tabular}
}}
\end{table}

Ultimately, the most conclusive evidence for the futility of the orange states comes from studying the stability of the $f_2$ kink under various \textit{spectral assumptions}. The idea is to re-run the positivity bounds of figure \ref{fig:kink-plot} making increasingly stronger assumptions about the allowed high-energy spectrum. If the $f_2$ kink is excluded, the assumptions were too strong. If, on the other hand, the kink survives, there exists \textit{at least one solution} which is compatible with our assumptions and saturates that bound. The first assumption that one might consider is that all states lie below the line going through the rho and $f_2$ mesons.\footnote{A similar assumption (dubbed ``maximal spin constraint'') was introduced in \cite{Haring:2023zwu} in the context of string theory amplitudes to impose the existence of a linear leading Regge trajectory.} The $f_2$ kink is compatible with this assumption, proving that all the states in the upper triangle of figure~\ref{fig:spectrum-at-kink-A} are indeed spurious. On the other hand, the bound in regions away from the kink becomes stronger, indicating that the extremal solution in other regions requires these states.\footnote{One can also see this by studying the couplings of the extremal solutions along the bound, comparing the couplings of the dominant trajectory to the other states. Doing this, one observes that the states away from the dominant trajectory have their smallest couplings precisely at the kink (shown in table \ref{tab:dominant-trajectory}).}

The new extremal spectrum at the $f_2$ kink obtained from \texttt{SDPB} is shown in figure~\ref{fig:spectrum-at-kink-B}. It still features the main (blue) trajectory of figure \ref{fig:spectrum-at-kink-A}, but it now contains a new collection of (orange) poles stretching between the curved trajectory and the linear cutoff. With more stringent assumptions, one can show that these are again spurious. These explorations then suggest that the main (blue) trajectory of figure \ref{fig:spectrum-at-kink-A} is actually the \textit{leading Regge trajectory} of the solution at the kink. Surprisingly, \texttt{SDPB} does not find any subleading (or daughter) trajectories below it. In fact, the $f_2$ kink is even consistent with spectral assumptions preventing any new poles to the right of this trajectory (at least for the first few spins), which would be necessary for daughter trajectories. This seems to imply that the $f_2$ kink may be saturated by an amplitude with a \textit{single (curved) Regge trajectory}, given by the blue states of figure \ref{fig:spectrum-at-kink}.

\subsection{Large $N$ QCD?}

We saw in section \ref{sec:Experiment}  that the cutoff $\widetilde M^2$ of the $f_2$ kink in figure \ref{fig:kink-plot} is compatible with the experimental value for the mass of the next spin-three meson, the $\rho_3$. We proceed now to compare the full extremal spectrum at the kink with the spectrum of real-world mesons. In figure \ref{fig:physical-spectrum} in the introduction, we have  plotted the first few resonances in the curved trajectory of the extremal spectrum against the experimentally-measured spectrum of leading mesons in pion-pion scattering, as reported in \cite{PDG}. Astonishingly, we find remarkably good agreement between the two spectra! All the physical states up to spin five match with the extremal spectrum at the kink to within experimental uncertainty.

This is clearly a major step forward in the quest for large $N$ QCD. By enforcing the exchange of a spin-two state, we have reached a new solution whose spectrum begins to show the main features of QCD. Far beyond a faint resemblance, it shows \textit{quantitative} agreement with real-world data; a remarkable fact, considering that we are working in the strict large $N$ limit. It is possible, of course, that such a near-perfect quantitative agreement might be a bit of a coincidence. 

A notable difference between our extremal solution and the expected spectrum of large $N$ QCD
is the \textit{absence of daughters}. As we discussed above, the extremal spectrum at the $f_2$ kink appears to contain a single Regge trajectory, in contrast with QCD, which is expected to contain an infinity of them. Nevertheless, subleading trajectories in real-world QCD appear to be significantly suppressed compared to the leading one \cite{PDG}. So perhaps it is not too far-fetched that this is truly a first glimpse of large $N$ QCD and that, upon cranking up the number of constraints, daughters will appear in our numerics.

From a more theoretical standpoint, 
 the hint of an extremal solution consisting of a single (curved) Regge trajectory is remarkable in itself, as one naively expects daughter trajectories to be required by crossing. One possibility is that a single Regge trajectory can be ``UV-completed'' by states at very high energy, in a similar spirit as the UV completion of a single spin-one exchange~\cite{Albert:2022oes}.
 Another logical possibility is that daughter trajectories kick in only at very high spin, escaping our numerical explorations. It would be very interesting to find explicit amplitudes with any of these properties.\footnote{Based on a holographic model, reference \cite{Katz:2005ir} proposed a spectrum with a leading trajectory that followed a square-root behavior. Our curved trajectory is significantly above that spectrum even for moderate spins.}

\subsection{EFT couplings}
\begin{figure}[h]
\centering
\includegraphics[width=0.95\textwidth]{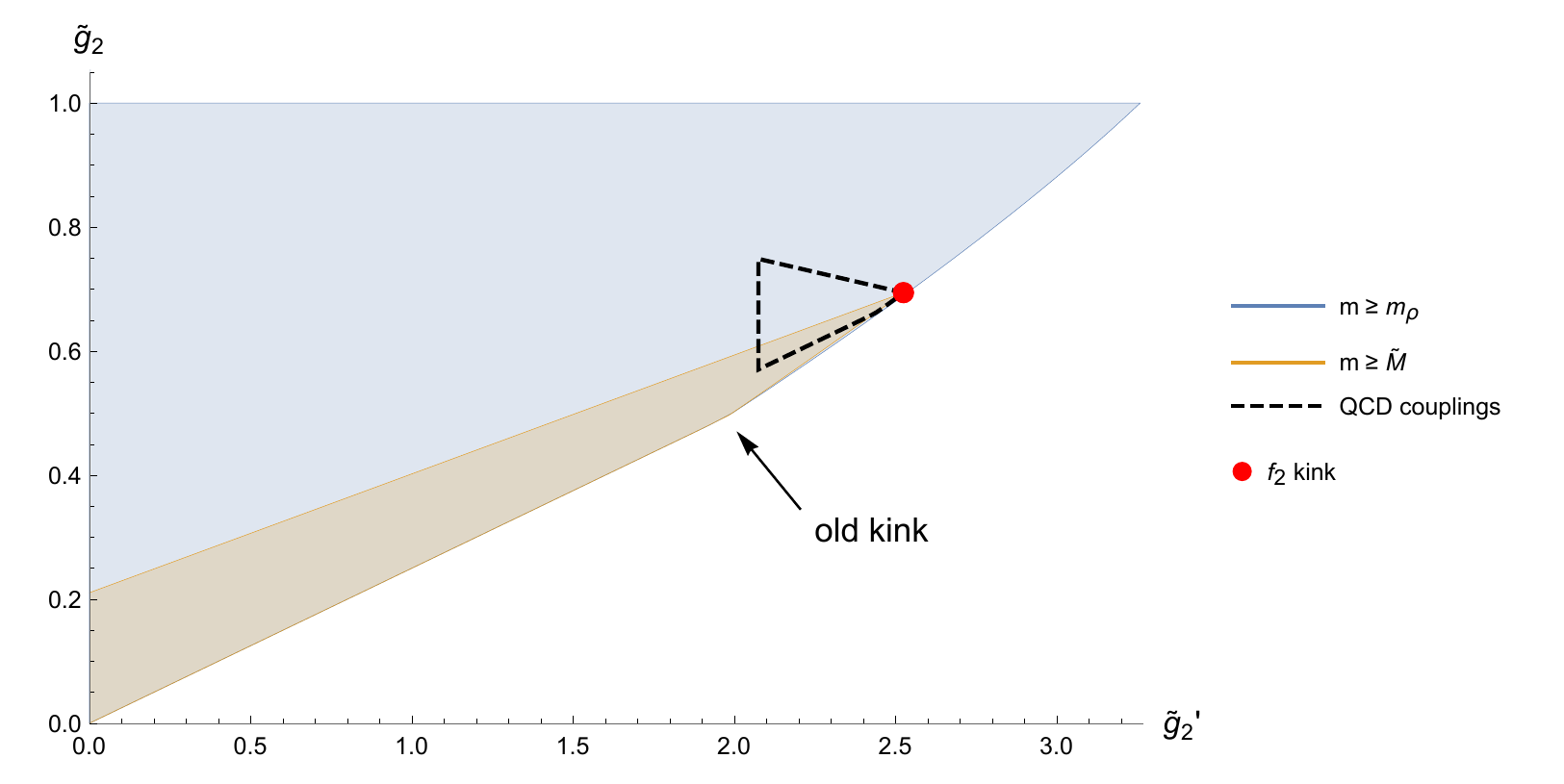}
\caption{
Allowed region in the space of (normalized) four-derivative EFT couplings $(\tilde g_2', \tilde g_2)$, produced with $n_{\mathrm{max}}=15$ under different assumptions.
Blue region: uniform cutoff $m\geq m_\rho$ as in \cite{Albert:2022oes}.
Orange region: cutoff $m\geq \widetilde M$ at the kink of figure \ref{fig:kink-plot}, and isolated rho and $f_2$ states with undetermined positive couplings.
Red dot: similar to the orange region but with on-shell couplings $(\tilde g_\rho^2, \tilde g_{f_2}^2)$ fixed from the kink in figure \ref{fig:couplings}.
Dashed black region: QCD values for $(\tilde g_\rho^2, \tilde g_{f_2}^2)$ and physical spectral assumptions \eqref{eq:physicalAssumptions}.
}\label{fig:EFTcouplings}
\end{figure}

We conclude by exploring where the extremal solution at the $f_2$
kink
lies inside the exclusion plot of \cite{Albert:2022oes} in the space of (normalized) four-derivative couplings  $(\tilde g_2',\tilde g_2)$. We do this by re-running the optimization problem of \cite{Albert:2022oes} with refined data from the new solution. This shrinks the allowed region, restricting to the EFTs compatible with a healthy UV completion which further satisfy the new assumptions. Assuming, first, isolated rho and $f_2$ states (with unfixed but positive couplings) below the cutoff $\widetilde M^2 \approx 4.748m_\rho^2$ at the kink of figure \ref{fig:kink-plot} produces the orange region of figure \ref{fig:EFTcouplings}. Further fixing the on-shell couplings to
\begin{equation}
    \tilde g_{\rho}^2\approx 0.611\,, \qquad \tilde g_{f_2}^2\approx  0.400\,,
\end{equation}
as read off from the tip in figure \ref{fig:couplings} shrinks the allowed region to almost a point. This is marked in figure \ref{fig:EFTcouplings} by a red dot.

\begin{figure}[h]
\centering
\includegraphics[width=0.65\textwidth]{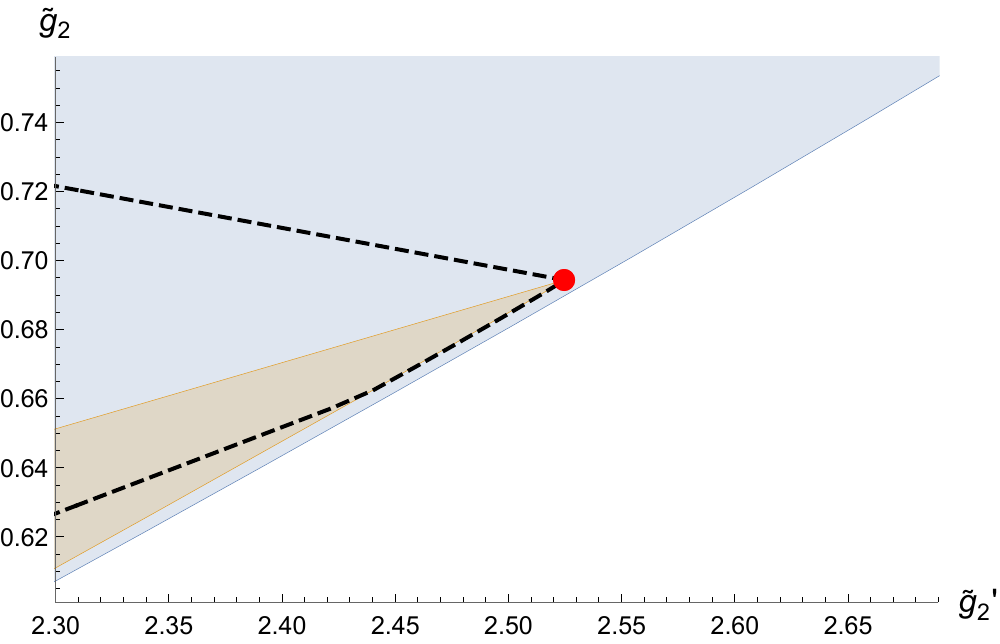}
\caption{
Zoomed-in version of figure~\ref{fig:EFTcouplings}.
}\label{fig:EFTcouplings-zoom}
\end{figure}

Surprisingly, we see that this solution lies close to the previous numerical bound, although as the zoomed-in version in figure~\ref{fig:EFTcouplings-zoom} shows, it does not quite saturate it. Perhaps it would if the numerical bound had fully converged.
This edge of the allowed blue region (stretching from the top-right corner to the kink discussed in \cite{Albert:2022oes}) is the only bound that still remains to be understood \cite{Albert:2022oes,Fernandez:2022kzi}. Perhaps our new solution -- with a single (curved) Regge trajectory -- might be the final piece of that puzzle, and trace the curved boundary as a function of $\frac{m^2_{f_2}}{m^2_\rho}$, but at present this is only a speculation.

For completeness, we carry out a similar analysis for the physical assumptions of real-world QCD. We include explicit rho and $f_2$ states with the couplings fixed to the physical values \eqref{eq:physicalCouplings}, and we take the following cutoffs: 
\begin{alignat}{2}\label{eq:physicalAssumptions}
     m&\geq m_\rho \qquad && \text{for}\quad J=0,1,\nonumber \\
     m&\geq m_{f_2} \qquad && \text{for}\quad J=2\,,\\
     m&\geq \widetilde M \qquad && \text{for}\quad J\geq 3\,.\nonumber
\end{alignat}
These spin-by-spin cutoffs are very conservative assumptions, accommodating for daughter trajectories starting anywhere after the rho. The positivity bounds with these physical assumptions produce the triangular-like island enclosed by dashed black lines in figure \ref{fig:EFTcouplings}. The fact that fixing the rho and $f_2$ couplings shrinks the allowed region to such a small island highlights the fact that low-energy couplings are mostly determined by the low-lying resonances, in line with the phenomenological success of old ideas like vector meson dominance \cite{sakurai}. The shape of the island is due to the freedom of adding low-lying states, such as scalars, for masses as low as $m_\rho$.

\section{Conclusions}\label{sec:conclusions}

In this work we have pushed the exploration of pion scattering in large $N$ QCD from a modern bootstrap perspective. Our starting point was a recognition that explicit amplitudes that saturate bounds on Wilson coefficients often lack intermediate states with spin $J > 1$ or, if present, exhibit unphysical characteristics violating locality. Recognizing the richer spectrum of resonances in QCD organized into Regge trajectories, we introduced a crucial assumption -- the existence of a spin-two intermediate state in pion scattering.

The geometry of the null constraints led to a series of spectral no-go theorems. A robust conclusion is that a single massive spin-two exchange cannot be reconciled by adding states at an arbitrarily high scale: a whole infinite tower of higher spin states with finite masses is necessary to reproduce the correct Regge behavior.

A natural strategy emerged from these insights: enforcing in the pion amplitude the presence of a $J=2$ massive state, in addition to the $J=1$ rho meson. This led to the discovery of a novel extremal solution characterized by a stable kink in the curve depicting the maximum allowed coupling of the $J=2$ state as a function of the cutoff scale $\widetilde{M}$. The exploration of this extremal solution is the principal achievement of our work.

Remarkably, our extremal solution exhibits a ratio of on-shell couplings $g_{\pi\pi f_2}/g_{\pi\pi \rho}$ that aligns perfectly with real-world values. What's more, the solution appears to trace a full Regge trajectory, and its low-lying states, especially for $J=3, 4, 5$, quantitatively match the experimental meson spectrum. This intriguing alignment raises the question of whether we have effectively cornered large $N$ QCD.
Upon closer scrutiny, we acknowledge the apparent sparsity of our solution's spectrum, which shows no evidence for daughter Regge trajectories. This prompts caution and invites further exploration.  

There are several natural directions for future work. Within the
dual bootstrap framework, an obvious next step is the study of the
complete mixed system of $2\to 2$ amplitudes with both pions and rhos as external states~\cite{rhos}. The spectral assumption that intermediate $J=2$ states 
must appear in this much richer system is likely to be very powerful. There are certain robust physical features of large $N$ QCD that are however difficult (though perhaps not impossible) to impose in a dual framework. One is the fact that Regge trajectories must be asymptotically linear~\cite{Caron-Huot:2016icg}. The other is that high-energy, fixed-angle scattering should be power-like (with logarithmic corrections), according to the predictions of asymptotic freedom. Perhaps a primal approach, along the lines of \cite{Veneziano:2017cks, Haring:2023zwu} may be better suited to impose these properties.

In conclusion, our exploration has brought us tantalizingly close to large $N$ QCD. Our novel extremal solution, with its intriguing kink, Regge trajectory and even quantitative  alignment with the real world, raises optimism but also underscores the need for further refinement.
The journey continues.

\section*{Acknowledgments}

We would like to thank Ilija Buri\'c, Andrea Guerrieri, Denis Karateev, Igor Klebanov, Waltraut Knop, Piotr Tourkine, Balt van Rees and Alexander Zhiboedov for interesting discussions and comments.
This work has received funding from the European Research Council (ERC) under the European Union's Horizon 2020 research and innovation program (grant agreement no.~758903). 
The work of JA and LR was supported in part
by the National Science Foundation under Grant No. NSF PHY-2210533. LR is supported
in part by Simons Foundation grants 397411 (Simons Collaboration on the Nonperturbative
Bootstrap) and 681267 (Simons Investigator Award). We thank the KITP, Santa Barbara,
for hospitality during the workshop “Bootstrapping quantum gravity”, when some of this
work was carried out. KITP workshops are supported in part by the National Science
Foundation under Grant No. NSF PHY-1748958. Numerical computations have been performed on the cluster of the Scientific Computing Center at INFN-PISA.

\appendix

\section{Real-world QCD}\label{app:RealWorld}
In this appendix we explain in detail the extraction of the rho and $f_2$ (normalized) on-shell couplings to real-world data. These are ultimately determined from their measured decay rates, available in \cite{PDG}, but there are some subtleties in the extraction worth pointing out. We first review the construction of meson exchange amplitudes using the framework of on-shell vertices, developed in \cite{Caron-Huot:2022jli}. This fixes the normalization of the interaction vertices, which we then use to compute meson decay rates. We end by plugging in experimental results for the latter, which determines the values of $g_{\pi\pi\rho}^2$ and $g_{\pi\pi f_2}^2$ in real-world QCD.

\subsection{Tree-level exchanges}
We start by discussing the tree-level exchange of a heavy meson $X$ in $2\to 2$ pion scattering. By unitarity, the corresponding amplitude must consist of a simple pole with a factorized residue between incoming and outgoing states. There are various ways to determine the residue. Perhaps the cleanest is to use the formalism of on-shell vertices described in \cite{Albert:2023jtd}, first introduced in \cite{Caron-Huot:2022jli}. The idea is to look for invariant tensors transforming in the different representations of the external legs in a three-point vertex, and then glue them together.

There is only one interaction vertex kinematically allowed between a massive meson $X$ and two pions, but its flavor part depends on whether the spin $J$ of $X$ is even or odd~\cite{Albert:2023jtd},
\begin{equation}\label{eq:vab}
    v_{ab}(n) \equiv k_J\, g_{\pi\pi X}\, n^{((\mu_1}n^{\mu_2}\cdots n^{\mu_{J}))}\times \begin{cases} d_{abc} & J=\text{even}\\
    f_{abc} & J=\text{odd}\,.
    \end{cases}
\end{equation}
The vertices are built out of a traceless-symmetric product of $J$ copies of the vector ${n^\mu \equiv p_2^\mu - p_1^\mu}$ involving the pion momenta. The flavor part consists of an invariant $U(N_f)$ tensor with three adjoint indices; $a,b$ for the pions and $c$ for $X$. The symmetry of the flavor tensor is linked to the parity of $J$ so that the full vertex remains invariant under the exchange of the two pions. The constant $g_{\pi\pi X}$ defines the \textit{on-shell coupling} between $X$ and two pions,\footnote{In terms of an effective Lagrangian, these vertices would be produced from interactions of the form
$$   \mathcal L_{\text{int}}\propto g_{\pi\pi X}\, \pi^a \partial^{\mu_1}\cdots \partial^{\mu_J}\pi^b \, X_{\mu_1\cdots \mu_{J}}^c \times \begin{cases} d_{abc} & J=\text{even}\\
    f_{abc} & J=\text{odd}\,,
    \end{cases}
$$
suitably normalized. We find it more convenient to define the couplings at the level of the on-shell vertices to avoid ambiguities due to integration by parts and field redefinitions.} and the factor $k_J$ is an arbitrary normalization constant which we will fix shortly.

After these preliminaries, we may write the $s$-channel contribution of an $X$ exchange to the full four-pion amplitude as
\begin{equation}
    {\mathcal T}_{abcd}^{s-\text{ch}} = \frac{1}{m_A^2 - s}\left( v_{cd}(n')^*, v_{ab}(n)\right)\,,
\end{equation}
where $n' \equiv p_3 - p_4$ and $(-,-)$ denotes a contraction of the flavor and spin indices of $X$. The kinematic part of this contraction can be evaluated using the rule from \cite{Caron-Huot:2022jli} (see also~\cite{Albert:2023jtd}) to contract traceless-symmetric products of $n^\mu$ in $D$ dimensions
\begin{equation}\label{eq:vcontraction}
    n'_{((\mu_1}n'_{\mu_2}\cdots n'_{\mu_{J}))} n^{((\mu_1}n^{\mu_2}\cdots n^{\mu_{J}))} = \frac{(D-3)_J}{2^J(\tfrac{D-3}{2})_J}|n'|^J|n|^J\legP_J\left(\frac{n\cdot n'}{|n||n'|}\right)\,.
\end{equation}
Noting that for massless pions $|n'|^2=|n|^2=m_X^2$ and $n\cdot n' = s + 2u$, we then find
\begin{equation}\label{eq:Ts-ch}
    {\mathcal T}_{abcd}^{s-\text{ch}}= \frac{1}{m_X^2 - s}
    k_J^2\, g_{\pi\pi X}^2
    \frac{(1)_J}{2^J(\tfrac{1}{2})_J}
    m_X^{2J} \legP_J\left(1+\frac{2u}{m_X^2}\right)
    \times \begin{cases} d\indices{_{ab}^e}d_{cde} & J=\text{even}\\
    f\indices{_{ab}^e}f_{cde} & J=\text{odd}\,,
    \end{cases} 
\end{equation}
up to analytic terms, which we ignore.

To extract the contribution of this exchange to the disk amplitude $M(s,u)$, we compare \eqref{eq:Ts-ch} to our parametrization \eqref{eq:Tabcd} of the four-pion amplitude using the identities
\begin{equation}
    d\indices{_{ab}^e}d_{cde} =\, 2\tr{\{T_a,T_b\}\{T_c,T_d\}}\,, \qquad
    f\indices{_{ab}^e}f_{cde} = -2\tr{[T_a,T_b][T_c,T_d]}  \,.
\end{equation}
It is then easy to see that, regardless of whether $J=\text{even}$ or odd, we get
\begin{equation}
    M_{s-\text{ch}}(s,u) = \frac{1}{2} \frac{1}{m_X^2 - s}
    k_J^2\, g_{\pi\pi X}^2
    \frac{(1)_J}{2^J(\tfrac{1}{2})_J}
    m_X^{2J} 
    \legP_J\left(1+\frac{2u}{m_X^2}\right)\,.
\end{equation}
We conclude that we must choose
\begin{equation}\label{eq:kJ}
    k_J = \left(\frac{(1)_J}{2^{J+1}(\tfrac{1}{2})_J}m_X^{J-1}\right)^{-\frac{1}{2}}\,,
\end{equation}
to recover the tree-level exchange amplitudes \eqref{eq:Mlow-exch} from the main text, which we reproduce here for convenience,
\begin{equation}
    M_{s-\text{ch}}(s,u) = g_{\pi\pi X}^2\frac{m_X^2}{m_X^2 - s}\legP_J\left(1+\frac{2u}{m_X^2}\right)\,.
\end{equation}

\subsection{Decay rates}
Having normalized the three-point vertices, we now turn to decay rates. We start from the usual formula for the decay rate of a heavy meson $X$ with polarization $\lambda$ and flavor index $c$ into two pions $\pi^a+\pi^b$,
\begin{equation}\label{eq:decayIntegral}
    \Gamma_\lambda^{abc} = 
    \frac{1}{2m_X}
    \int\frac{d^3 \vec{p}_1}{(2\pi)^3 2E_1} \int\frac{d^3 \vec{p}_2}{(2\pi)^32 E_2}
    \left|M(X_\lambda^c\to \pi^a+\pi^b)\right|^2 (2\pi)^4 \delta^{(4)}(p_1+p_2+p_3)\,.
\end{equation}
The total decay rate of $X$ into two pions is then given by the sum over outgoing flavor and the average of the incoming flavor and polarizations,
\begin{equation}
    \Gamma(X\to \pi + \pi) = \frac{1}{2}\frac{1}{(2J+1)} \sum_\lambda \frac{1}{\text{dim}\,\mathcal R}\sum_{a,b,c}\Gamma_\lambda^{abc}\,.
\end{equation}
Here $J$ and $\mathcal R$ are respectively the spin and flavor representation of $X$. The factor of $\frac{1}{2}$ is to avoid overcounting in the sum over outgoing identical states.

The key point that connects to the previous discussion is that the absolute square of the amplitude summed over the quantum numbers of $X$ can also be expressed in terms of on-shell three-point vertices. Namely,
\begin{equation}
    \sum_\lambda \sum_c |M(X_\lambda^c\to \pi^a+\pi^b)|^2 = \left( v_{ab}(n)^*, v_{ab}(n)\right)\,.
\end{equation}
This contraction is immediate to evaluate using again \eqref{eq:vcontraction}. Taking into account the normalization \eqref{eq:kJ} and performing the sum over outgoing flavor, we get
\begin{equation}
    \sum_\lambda \sum_{a,b,c} |M(X_\lambda^c\to \pi^a+\pi^b)|^2 
    = 2 g_{\pi\pi X}^2 \frac{|n|^{2J}}{(m_X^2)^{J-1}} \times \begin{cases} d_{abc}d_{abc} & J=\text{even}\\
    f_{abc}f_{abc} & J=\text{odd}\,,
    \end{cases} 
\end{equation}
where summation over repeated indices is understood.

Keeping now a nonzero mass for the pion (which is meaningful in real-world decays), we have that $|n|^2 = (p_2-p_1)^2 = m_X^2 - 4m_\pi^2$. Evaluating the kinematic integral \eqref{eq:decayIntegral} we find the final expression for the total decay rate in terms of the on-shell coupling $g_{\pi\pi X}^2$,
\begin{equation}\label{eq:fullGamma}
    \Gamma(X\to \pi + \pi) = \frac{1}{2}\frac{1}{(2J+1)} \frac{1}{\text{dim}\,\mathcal R}
    \frac{g_{\pi\pi X}^2 m_X}{8\pi}\left(1 - \frac{4m_\pi^2}{m_X^2}\right)^{J+\frac{1}{2}\hspace{-5 pt}} \times \begin{cases} d_{abc}d_{abc} & J=\text{even}\\
    f_{abc}f_{abc} & J=\text{odd}\,.
    \end{cases}
\end{equation}
Since the decay rate $\Gamma$ has dimensions of mass, we see that our normalization conventions are such that the couplings $g_{\pi\pi X}$ are dimensionless. It only remains to evaluate the flavor factor, which we will do in the next section.

\subsection{Physical mesons}
We are finally ready to determine the physical values for the on-shell meson couplings from experimental data. We first discuss the rho meson, and then turn to the $f_2$.

\subsubsection*{The rho meson}
Here and onward we restrict to $N_f=2$ mesons. For the rho, a spin-one meson in the adjoint of $SU(2)$ (i.e.\ isospin $I=1$), we have $\text{dim}\,\mathcal R = 3$ and $f_{abc}f_{abc} = N_f(N_f^2-1) = 6$. Plugging this into \eqref{eq:fullGamma} yields
\begin{equation}
    \Gamma(\rho\to \pi + \pi) = \frac{g_{\pi\pi \rho}^2}{24 \pi} m_\rho
    \left(1 - \frac{4m_\pi^2}{m_\rho^2}\right)^{\frac{3}{2}\hspace{-5 pt}}\,.
\end{equation}
This only differs from the result quoted in \cite{Albert:2022oes} by a factor of $1/2$, which stems from our new normalization for the on-shell coupling, c.f.\ \eqref{eq:Mlow-exch}.

Using \cite{PDG} $m_{\rho} = 775.26 \pm 0.23\,$MeV, $m_{\pi^{\pm}} = 139.57039 \pm 0.00018\,$MeV and $\Gamma(\rho\to 2\pi) = 149.1 \pm 0.8\,$MeV, we obtain
\begin{equation}
    g_{\pi\pi\rho}^2 = 17.86 \pm 0.10\,.
\end{equation}
In turn, using that $g_{1,0} = \frac{1}{2f_\pi^2}$ with $f_\pi = 92.1\pm 0.8\,$MeV, we obtain for the normalized coupling
\begin{equation}
    \tilde g_\rho^2 \equiv \frac{g_{\pi\pi\rho}^2}{g_{1,0}m_\rho^2} \simeq 0.504 \pm 0.009\,.
\end{equation}

\subsubsection*{The $f_2$ meson}
We now turn to the $f_2$ meson; a spin-two meson in the trivial representation of $SU(2)$. The restriction from the $U(N_f)$ adjoint representation to the singlet plus adjoint of $SU(N_f)$ is done by writing explicitly the generator of $\mathfrak u(N_f)$ proportional to the identity; $T_0 = \frac{1}{\sqrt{2N_f}}\mathbb 1$. For the singlet of $SU(N_f)$, the $d$-symbol in \eqref{eq:vab} becomes
\begin{equation}
    d_{ab0} = 2\text{Tr}\Big(\{ T_a,T_b\}\frac{\mathbb 1}{\sqrt{2N_f}}\Big) = \sqrt{\frac{2}{N_f}}\delta_{ab}\,,
\end{equation}
where $a,b$ are now $SU(N_f)$ adjoint indices, and so we have
\begin{equation}
    d_{abc}d_{abc} = \frac{2}{N_f} (N_f^2-1) = 3\,.
\end{equation}
Since the dimension of the trivial representation is $\text{dim}\, \mathcal R = 1$, we find
\begin{equation}
    \Gamma(f_2\to \pi + \pi) = \frac{3 g_{\pi\pi f_2}^2}{80 \pi} m_{f_2}
    \left(1 - \frac{4m_\pi^2}{m_{f_2}^2}\right)^{\frac{5}{2}\hspace{-5 pt}}\,.
\end{equation}

Using \cite{PDG} $m_{f_2} = 1275.5 \pm 0.8\,$MeV and $\Gamma(f_2\to 2\pi) = 157^{+6}_{-2}\,$MeV, we obtain
\begin{equation}
    g_{\pi\pi f_2}^2 = 11.7 ^{+0.4}_{-0.1}\,.
\end{equation}
The corresponding normalized ratio is then
\begin{equation}
    \tilde g_{f_2}^2 \equiv \frac{g_{\pi\pi f_2}^2}{g_{1,0}m_\rho^2} \simeq 0.329^{+0.013}_{-0.007}\,.
\end{equation}

\section{Extremal spectra along the bound}

\label{app:alongbound}
In this appendix we report the results from studying the extremal spectrum along the exclusion boundary where  the $f_2$ coupling is maximized, figure~\ref{fig:kink-plot} in the main text. At each value of $m_\rho^2/\widetilde M^2$, we extract the spectrum using the extremal functional method with \texttt{spectrum.py} \cite{Komargodski:2016auf,Simmons-Duffin:2016wlq}. Then for each spin $J=3,4,5,6,7$ we select the state with largest coupling $\tilde g_X^2$. At the $f_2$ kink, this selects the states on the dominant trajectory, shown in blue in figure~\ref{fig:spectrum-at-kink}.

\begin{figure}[h]
\centering
\includegraphics[width=0.82\textwidth]{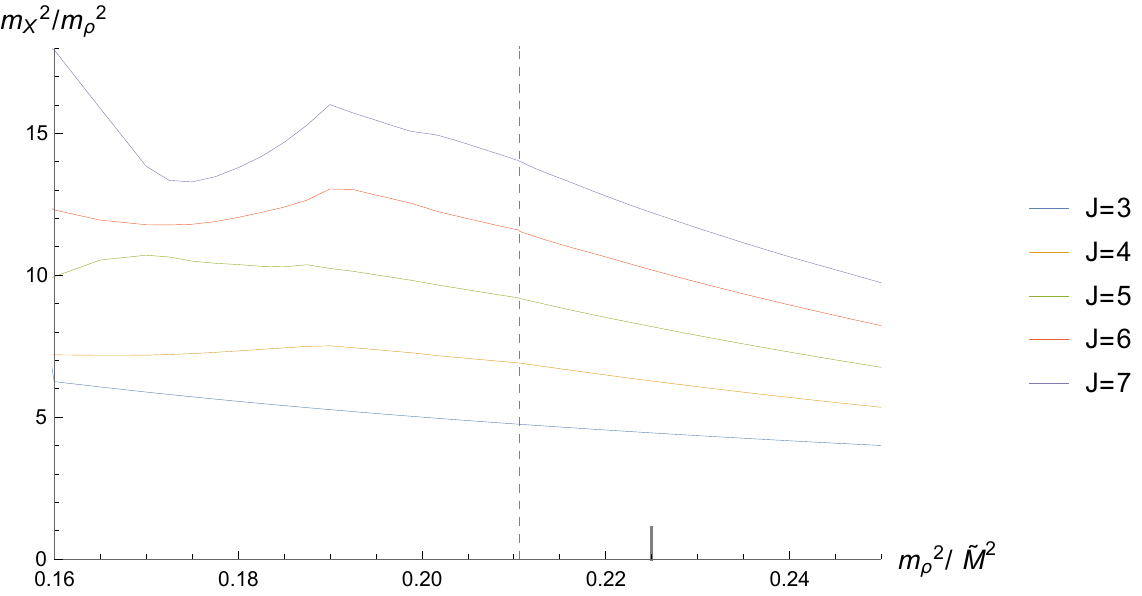}
\caption{The masses on the dominant trajectory along the bound. The kink is marked by the dashed vertical line. The value where $\widetilde M$ intersects the linear extrapolation through spins 1 and 2 is marked with the short gray bar. The plot was made with $n_{\mathrm{max}}=17$, with a denser sampling of $\widetilde M$ values near the kink.
}\label{fig:spectrum-at-bound}
\end{figure}

\begin{figure}[h]
\centering
\includegraphics[width=0.82\textwidth]{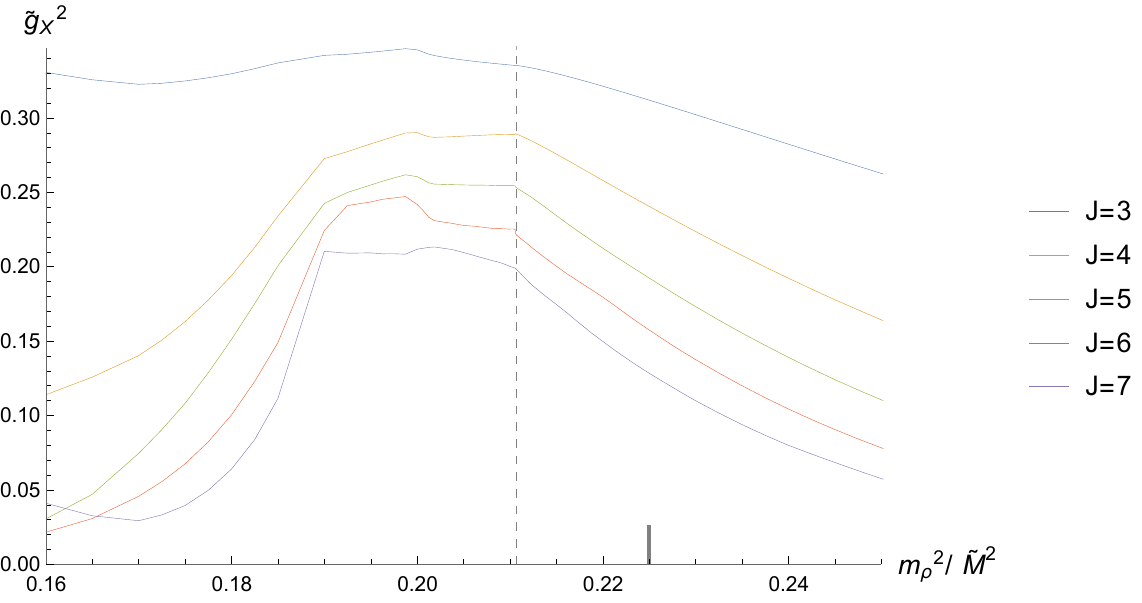}
\caption{The (normalized) on-shell couplings of the dominant trajectory along the bound. The kink is marked by the dashed vertical line, the linear trajectory by the short gray bar. The plot was made with $n_{\mathrm{max}}=17$.
}\label{fig:coupling-at-bound}
\end{figure}

In figure~\ref{fig:spectrum-at-bound}, we plot the mass of the state with the dominant coupling at every spin, as we move along the bound of figure \ref{fig:kink-plot}. There is no noticeable feature when we cross the point corresponding to the $f_2$ kink (dashed vertical line).
For the couplings, on the other hand, figure~\ref{fig:coupling-at-bound} shows a clear feature at the kink, where the dependence of the couplings as a function of $m_\rho^2/\widetilde M^2$ has a discontinuous first derivative. 
Note also that in both plots, nothing special happens when we pass the point of a linear trajectory, marked by the short gray line at $m_\rho^2/\widetilde M^2\approx 0.225$.

\section{Variations of the Lovelace--Shapiro amplitude}\label{app:LS}
In this appendix we discuss some variations of the Lovelace--Shapiro (LS) amplitude; an analytic amplitude classically used to model pion scattering at large $N$ \cite{Lovelace:1968kjy,Shapiro:1969km} (see also~\cite{Bianchi:2020cfc} for a recent discussion),
\begin{equation}\label{eq:LSamplitude}
    M_{\mathrm{LS}}(s,u)=-\frac{\Gamma(1-\alpha(s))\Gamma(1-\alpha(u))}{\Gamma(1-\alpha(s)-\alpha(u))}\,, \quad \text{with} \quad \alpha(s)\equiv \alpha_0 + \alpha' s\,.
\end{equation}
This amplitude  (for a suitably chosen Regge trajectory $\alpha(s)$) satisfies all the constraints of our problem. In contrast with the more familiar Veneziano amplitude, it contains a spin-one resonance in the first pole, which can be interpreted as the rho meson. In order to satisfy the ``Adler zero constraint'' (i.e.\ the vanishing of the amplitude as $s,u\to 0$), one usually chooses $\alpha_0 = 1/2$. Since our bootstrap setup is blind to this constraint,\footnote{This constraint imposes the vanishing of the constant term in the low-energy expansion \eqref{eq:Mlow}, but this piece cannot be accessed via dispersion relations without assuming an unphysical Regge behavior. See section 7.1 in \cite{Albert:2023jtd} for a discussion of more general \textit{Goldstone constraints} in the context of pion-photon scattering.} however, we can consider more general trajectories.

We fix the trajectory by requiring that it goes through the rho and $f_2$ mesons, i.e.\ $\alpha(m_\rho^2)=1$ and $\alpha(m_{f_2}^2)=2$, which gives
\begin{equation}
    \alpha_0 = 1-\frac{m_\rho^2}{m_{f_2}^2-m_\rho^2}\,, \qquad \alpha' = \frac1{m_{f_2}^2-m_\rho^2}\,.
\end{equation}
This produces a family of generalized LS amplitudes parametrized by the ratio $m_{f_2}^2/m_\rho^2$. For $m_{f_2}^2/m_\rho^2 = 3$, we get back the usual LS amplitude. The spectrum of this amplitude arranges in \textit{linear} Regge trajectories, with a leading one going through the rho and the $f_2$, and subleading trajectories at the same masses but lower spins.

This amplitude, however, is not unitary unless $m_{f_2}^2/m_{\rho}^2\geq 3$. But there is a range
\begin{equation}
    5/2<m_{f_2}^2/m_{\rho}^2<3,
\end{equation}
where the only negativity comes from the scalar state degenerate with the $f_2$ (and also the scalar at mass $3m_{f_2}^2-2m_\rho^2$ in a smaller subregion). Fortunately, the physical value \eqref{eq:PhysicalValue} is in this range.
Since scalar exchanges are allowed by our assumptions on their own (recall from \eqref{eq:NC} that they trivially satisfy all null constraints),
this means that we can compensate this negativity simply by adding a scalar exchange with a suitable coefficient. Schematically, 
\begin{align}
\label{eq:unit-LS}
    M_{\text{unit-LS}}(s,u)&=M_{\text{LS}}(s,u)-\kappa\left(
    \frac{m^2_{f_2}}{m^2_{f_2}-s}+\frac{m^2_{f_2}}{m^2_{f_2}-u}\right)\,,
\end{align}
where $\kappa$ is the (negative) scalar coupling in $M_{\text{LS}}(s,u)$.
This gives a unitary amplitude consistent with all of our assumptions.\footnote{As is usually the case for string-like amplitudes, there is no general proof of unitarity for all residues. We have checked that the partial wave expansion remains positive for the residues of the first 150 poles.} Its corresponding couplings are,
\begin{equation}
    \tilde g_\rho^2 = 0.42502\,, \qquad 
    \tilde g_{f_2}^2 = 0.22392\,,
\end{equation}
and, since the trajectory is linear, the first spin-three exchange sits at
\begin{equation}
    \frac{\widetilde M^2}{m_\rho^2} = 2\frac{m_{f_2}^2}{m_\rho^2} - 1 = 4.445\,.
\end{equation}

The fact that in this system we are always free to add or subtract scalars was exploited in \cite{Fernandez:2022kzi} to propose a spin-0 subtracted version of the LS amplitude, where \textit{all scalar states} are removed by hand. It was shown there that this amplitude lies much closer than the standard LS amplitude to the positivity bounds derived in \cite{Albert:2022oes}. Here, we can proceed similarly and remove all scalar states from \eqref{eq:LSamplitude}. Namely,
\begin{equation}
\label{eq:LS-scalars}
    M_{\mathrm{LS}_{-0}}(s,u)=M_{\mathrm{LS}}(s,u)-\sum_{n=1}^\infty\kappa_{n}\left(
    \frac{m^2_n}{m_n^2-s}+\frac{m^2_n}{m^2_n-u}
    \right)\,,
\end{equation}
where $m_n^2 = \frac{1}{\alpha'}(n-\alpha_0)$, and $\kappa_n$ are the scalar couplings in $M_{\text{LS}}(s,u)$. This subtraction does not change the on-shell couplings $g_{\pi\pi\rho}^2,g_{\pi\pi f_2}^2$, but it does make the EFT coupling $g_{1,0}$ smaller, increasing the normalized ratios \eqref{eq:gtilde}. We now find\footnote{In practice, we have not been able to perform the infinite sum in closed form. We subtracted scalars up to high order and then extrapolated using a quadratic fit.
}
\begin{equation}
    \tilde g_\rho^2 = 0.52748\,, \qquad 
    \tilde g_{f_2}^2 = 0.27790\,.
    \label{eq:Linear-minus-scalars}
\end{equation}
Since the leading trajectory is not changed by the subtraction, $\widetilde M^2$ remains unchanged.
We have marked the location of this amplitude by a gray dot in figure \ref{fig:kink-plot}. We see that it is well within the bounds, as it should be.

Apart from removing scalars, one can consider linear combinations with more general string amplitudes such as LS amplitudes with different slopes and intercepts, but still satisfying the correct Regge behavior. It would be interesting to explore the space of amplitudes ruled-in by such an ansatz, in the spirit of \cite{Veneziano:2017cks,Haring:2023zwu}.\footnote{One has to be careful with the ansatz: as observed in \cite{Haring:2023zwu}, certain finite linear combinations of LS amplitudes necessarily violate unitarity.}

\bibliographystyle{ytphys}
\bibliography{references}

\providecommand{\href}[2]{#2}\begingroup\raggedright\begin{thebibliography}{10}

\bibitem{PDG}
{\bfseries Particle Data Group} Collaboration, R.~L. Workman {\em et al.},
  {\slshape {Review of Particle Physics},}
  \href{http://dx.doi.org/10.1093/ptep/ptac097}{{\em PTEP} {\bfseries 2022}
  (2022) 083C01}.

\bibitem{Albert:2022oes}
J.~Albert and L.~Rastelli, {\slshape {Bootstrapping pions at large N},}
  \href{http://dx.doi.org/10.1007/JHEP08(2022)151}{{\em JHEP} {\bfseries 08}
  (2022) 151}, \href{http://arxiv.org/abs/2203.11950}{{
  arXiv:2203.11950~[hep-th]}}.

\bibitem{Fernandez:2022kzi}
C.~Fernandez, A.~Pomarol, F.~Riva, and F.~Sciotti, {\slshape {Cornering
  large-N$_{c}$ QCD with positivity bounds},}
  \href{http://dx.doi.org/10.1007/JHEP06(2023)094}{{\em JHEP} {\bfseries 06}
  (2023) 094}, \href{http://arxiv.org/abs/2211.12488}{{
  arXiv:2211.12488~[hep-th]}}.

\bibitem{Albert:2023jtd}
J.~Albert and L.~Rastelli, {\slshape {Bootstrapping Pions at Large $N$. Part
  II: Background Gauge Fields and the Chiral Anomaly},}
  \href{http://arxiv.org/abs/2307.01246}{{ arXiv:2307.01246~[hep-th]}}.

\bibitem{Ma:2023vgc}
T.~Ma, A.~Pomarol, and F.~Sciotti, {\slshape {Bootstrapping the chiral anomaly
  at large N$_{c}$},} \href{http://dx.doi.org/10.1007/JHEP11(2023)176}{{\em
  JHEP} {\bfseries 11} (2023) 176}, \href{http://arxiv.org/abs/2307.04729}{{
  arXiv:2307.04729~[hep-th]}}.

\bibitem{Li:2023qzs}
Y.-Z. Li, {\slshape {Effective field theory bootstrap, large-N $\chi$PT and
  holographic QCD},} \href{http://arxiv.org/abs/2310.09698}{{
  arXiv:2310.09698~[hep-th]}}.

\bibitem{Pham:1985cr}
T.~N. Pham and T.~N. Truong, {\slshape {Evaluation of the derivative quartic
  terms of the meson chiral Lagrangian from forward dispersion relation},}
  \href{http://dx.doi.org/10.1103/PhysRevD.31.3027}{{\em Phys. Rev. D}
  {\bfseries 31} (1985) 3027}.

\bibitem{Pennington:1994kc}
M.~R. Pennington and J.~Portoles, {\slshape {The Chiral Lagrangian parameters,
  $\bar\ell_1$, $\bar\ell_2$, are determined by the $\rho$-resonance},}
  \href{http://dx.doi.org/10.1016/0370-2693(94)01551-M}{{\em Phys. Lett. B}
  {\bfseries 344} (1995) 399--406},
  \href{http://arxiv.org/abs/hep-ph/9409426}{{ arXiv:hep-ph/9409426}}.

\bibitem{Ananthanarayan:1994hf}
B.~Ananthanarayan, D.~Toublan, and G.~Wanders, {\slshape {Consistency of the
  chiral pion pion scattering amplitudes with axiomatic constraints},}
  \href{http://dx.doi.org/10.1103/PhysRevD.51.1093}{{\em Phys. Rev. D}
  {\bfseries 51} (1995) 1093--1100},
  \href{http://arxiv.org/abs/hep-ph/9410302}{{ arXiv:hep-ph/9410302}}.

\bibitem{Comellas:1995hq}
J.~Comellas, J.~I. Latorre, and J.~Taron, {\slshape {Constraints on chiral
  perturbation theory parameters from QCD inequalities},}
  \href{http://dx.doi.org/10.1016/0370-2693(95)01110-C}{{\em Phys. Lett. B}
  {\bfseries 360} (1995) 109--116},
  \href{http://arxiv.org/abs/hep-ph/9507258}{{ arXiv:hep-ph/9507258}}.

\bibitem{Dita:1998mh}
P.~Dita, {\slshape {Positivity constraints on chiral perturbation theory pion
  pion scattering amplitudes},}
  \href{http://dx.doi.org/10.1103/PhysRevD.59.094007}{{\em Phys. Rev. D}
  {\bfseries 59} (1999) 094007}, \href{http://arxiv.org/abs/hep-ph/9809568}{{
  arXiv:hep-ph/9809568}}.

\bibitem{Adams:2006sv}
A.~Adams, N.~Arkani-Hamed, S.~Dubovsky, A.~Nicolis, and R.~Rattazzi, {\slshape
  {Causality, analyticity and an IR obstruction to UV completion},}
  \href{http://dx.doi.org/10.1088/1126-6708/2006/10/014}{{\em JHEP} {\bfseries
  10} (2006) 014}, \href{http://arxiv.org/abs/hep-th/0602178}{{
  arXiv:hep-th/0602178}}.

\bibitem{Arkani-Hamed:2020blm}
N.~Arkani-Hamed, T.-C. Huang, and Y.-T. Huang, {\slshape {The EFT-hedron},}
  \href{http://dx.doi.org/10.1007/JHEP05(2021)259}{{\em JHEP} {\bfseries 05}
  (2021) 259}, \href{http://arxiv.org/abs/2012.15849}{{
  arXiv:2012.15849~[hep-th]}}.

\bibitem{Bellazzini:2020cot}
B.~Bellazzini, J.~Elias~Mir\'o, R.~Rattazzi, M.~Riembau, and F.~Riva, {\slshape
  {Positive moments for scattering amplitudes},}
  \href{http://dx.doi.org/10.1103/PhysRevD.104.036006}{{\em Phys. Rev. D}
  {\bfseries 104} (2021) 036006}, \href{http://arxiv.org/abs/2011.00037}{{
  arXiv:2011.00037~[hep-th]}}.

\bibitem{Tolley:2020gtv}
A.~J. Tolley, Z.-Y. Wang, and S.-Y. Zhou, {\slshape {New positivity bounds from
  full crossing symmetry},}
  \href{http://dx.doi.org/10.1007/JHEP05(2021)255}{{\em JHEP} {\bfseries 05}
  (2021) 255}, \href{http://arxiv.org/abs/2011.02400}{{
  arXiv:2011.02400~[hep-th]}}.

\bibitem{Caron-Huot:2020cmc}
S.~Caron-Huot and V.~Van~Duong, {\slshape {Extremal Effective Field Theories},}
  \href{http://dx.doi.org/10.1007/JHEP05(2021)280}{{\em JHEP} {\bfseries 05}
  (2021) 280}, \href{http://arxiv.org/abs/2011.02957}{{
  arXiv:2011.02957~[hep-th]}}.

\bibitem{Rattazzi:2008pe}
R.~Rattazzi, V.~S. Rychkov, E.~Tonni, and A.~Vichi, {\slshape {Bounding scalar
  operator dimensions in 4D CFT},}
  \href{http://dx.doi.org/10.1088/1126-6708/2008/12/031}{{\em JHEP} {\bfseries
  12} (2008) 031}, \href{http://arxiv.org/abs/0807.0004}{{
  arXiv:0807.0004~[hep-th]}}.

\bibitem{Poland:2011ey}
D.~Poland, D.~Simmons-Duffin, and A.~Vichi, {\slshape {Carving Out the Space of
  4D CFTs},} \href{http://dx.doi.org/10.1007/JHEP05(2012)110}{{\em JHEP}
  {\bfseries 05} (2012) 110}, \href{http://arxiv.org/abs/1109.5176}{{
  arXiv:1109.5176~[hep-th]}}.

\bibitem{Poland:2018epd}
D.~Poland, S.~Rychkov, and A.~Vichi, {\slshape {The Conformal Bootstrap:
  Theory, Numerical Techniques, and Applications},}
  \href{http://dx.doi.org/10.1103/RevModPhys.91.015002}{{\em Rev. Mod. Phys.}
  {\bfseries 91} (2019) 015002}, \href{http://arxiv.org/abs/1805.04405}{{
  arXiv:1805.04405~[hep-th]}}.

\bibitem{Rychkov:2023wsd}
S.~Rychkov and N.~Su, {\slshape {New Developments in the Numerical Conformal
  Bootstrap},} \href{http://arxiv.org/abs/2311.15844}{{
  arXiv:2311.15844~[hep-th]}}.

\bibitem{Lovelace:1968kjy}
C.~Lovelace, {\slshape {A novel application of regge trajectories},}
  \href{http://dx.doi.org/10.1016/0370-2693(68)90255-4}{{\em Phys. Lett. B}
  {\bfseries 28} (1968) 264--268}.

\bibitem{Shapiro:1969km}
J.~A. Shapiro, {\slshape {Narrow-resonance model with regge behavior for pi pi
  scattering},} \href{http://dx.doi.org/10.1103/PhysRev.179.1345}{{\em Phys.
  Rev.} {\bfseries 179} (1969) 1345--1353}.

\bibitem{EliasMiro:2022xaa}
J.~Elias~Miro, A.~Guerrieri, and M.~A. Gumus, {\slshape {Bridging positivity
  and S-matrix bootstrap bounds},}
  \href{http://dx.doi.org/10.1007/JHEP05(2023)001}{{\em JHEP} {\bfseries 05}
  (2023) 001}, \href{http://arxiv.org/abs/2210.01502}{{
  arXiv:2210.01502~[hep-th]}}.

\bibitem{Paulos:2016fap}
M.~F. Paulos, J.~Penedones, J.~Toledo, B.~C. van Rees, and P.~Vieira, {\slshape
  {The S-matrix bootstrap. Part I: QFT in AdS},}
  \href{http://dx.doi.org/10.1007/JHEP11(2017)133}{{\em JHEP} {\bfseries 11}
  (2017) 133}, \href{http://arxiv.org/abs/1607.06109}{{
  arXiv:1607.06109~[hep-th]}}.

\bibitem{Paulos:2016but}
M.~F. Paulos, J.~Penedones, J.~Toledo, B.~C. van Rees, and P.~Vieira, {\slshape
  {The S-matrix bootstrap II: Two dimensional amplitudes},}
  \href{http://dx.doi.org/10.1007/JHEP11(2017)143}{{\em JHEP} {\bfseries 11}
  (2017) 143}, \href{http://arxiv.org/abs/1607.06110}{{
  arXiv:1607.06110~[hep-th]}}.

\bibitem{Paulos:2017fhb}
M.~F. Paulos, J.~Penedones, J.~Toledo, B.~C. van Rees, and P.~Vieira, {\slshape
  {The S-matrix bootstrap. Part III: Higher dimensional amplitudes},}
  \href{http://dx.doi.org/10.1007/JHEP12(2019)040}{{\em JHEP} {\bfseries 12}
  (2019) 040}, \href{http://arxiv.org/abs/1708.06765}{{
  arXiv:1708.06765~[hep-th]}}.

\bibitem{Guerrieri:2021ivu}
A.~Guerrieri, J.~Penedones, and P.~Vieira, {\slshape {Where is string theory in
  the space of scattering amplitudes?},}
  \href{http://dx.doi.org/10.1103/PhysRevLett.127.081601}{{\em Phys. Rev.
  Lett.} {\bfseries 127} (2021) 081601},
  \href{http://arxiv.org/abs/2102.02847}{{ arXiv:2102.02847~[hep-th]}}.

\bibitem{Karateev:2022jdb}
D.~Karateev, J.~Marucha, J.~a. Penedones, and B.~Sahoo, {\slshape
  {Bootstrapping the a-anomaly in 4d QFTs},}
  \href{http://dx.doi.org/10.1007/JHEP12(2022)136}{{\em JHEP} {\bfseries 12}
  (2022) 136}, \href{http://arxiv.org/abs/2204.01786}{{
  arXiv:2204.01786~[hep-th]}}.

\bibitem{Haring:2022sdp}
K.~H\"aring, A.~Hebbar, D.~Karateev, M.~Meineri, and J.~Penedones, {\slshape
  {Bounds on photon scattering},} \href{http://arxiv.org/abs/2211.05795}{{
  arXiv:2211.05795~[hep-th]}}.

\bibitem{Miro:2023bon}
J.~E. Miro, A.~Guerrieri, and M.~A. Gumus, {\slshape {Extremal Higgs
  couplings},} \href{http://arxiv.org/abs/2311.09283}{{
  arXiv:2311.09283~[hep-ph]}}.

\bibitem{Guerrieri:2022sod}
A.~Guerrieri, H.~Murali, J.~Penedones, and P.~Vieira, {\slshape {Where is
  M-theory in the space of scattering amplitudes?},}
  \href{http://dx.doi.org/10.1007/JHEP06(2023)064}{{\em JHEP} {\bfseries 06}
  (2023) 064}, \href{http://arxiv.org/abs/2212.00151}{{
  arXiv:2212.00151~[hep-th]}}.

\bibitem{Guerrieri:2020bto}
A.~L. Guerrieri, J.~Penedones, and P.~Vieira, {\slshape {S-matrix bootstrap for
  effective field theories: massless pions},}
  \href{http://dx.doi.org/10.1007/JHEP06(2021)088}{{\em JHEP} {\bfseries 06}
  (2021) 088}, \href{http://arxiv.org/abs/2011.02802}{{
  arXiv:2011.02802~[hep-th]}}.

\bibitem{Guerrieri:2018uew}
A.~L. Guerrieri, J.~Penedones, and P.~Vieira, {\slshape {Bootstrapping QCD
  Using Pion Scattering Amplitudes},}
  \href{http://dx.doi.org/10.1103/PhysRevLett.122.241604}{{\em Phys. Rev.
  Lett.} {\bfseries 122} (2019) 241604},
  \href{http://arxiv.org/abs/1810.12849}{{ arXiv:1810.12849~[hep-th]}}.

\bibitem{He:2023lyy}
Y.~He and M.~Kruczenski, {\slshape {Bootstrapping gauge theories},}
  \href{http://arxiv.org/abs/2309.12402}{{ arXiv:2309.12402~[hep-th]}}.

\bibitem{Guerrieri:2023qbg}
A.~L. Guerrieri, A.~Hebbar, and B.~C. van Rees, {\slshape {Constraining
  Glueball Couplings},} \href{http://arxiv.org/abs/2312.00127}{{
  arXiv:2312.00127~[hep-th]}}.

\bibitem{EliasMiro:2021nul}
J.~Elias~Mir\'o and A.~Guerrieri, {\slshape {Dual EFT bootstrap: QCD flux
  tubes},} \href{http://dx.doi.org/10.1007/JHEP10(2021)126}{{\em JHEP}
  {\bfseries 10} (2021) 126}, \href{http://arxiv.org/abs/2106.07957}{{
  arXiv:2106.07957~[hep-th]}}.

\bibitem{Froissart:1961ux}
M.~Froissart, {\slshape {Asymptotic behavior and subtractions in the Mandelstam
  representation},} \href{http://dx.doi.org/10.1103/PhysRev.123.1053}{{\em
  Phys. Rev.} {\bfseries 123} (Aug, 1961) 1053--1057}.

\bibitem{Martin:1965jj}
A.~Martin, {\slshape {Extension of the axiomatic analyticity domain of
  scattering amplitudes by unitarity. 1.},}
  \href{http://dx.doi.org/10.1007/BF02720568}{{\em Nuovo Cim. A} {\bfseries 42}
  (1965) 930--953}.

\bibitem{Camanho:2014apa}
X.~O. Camanho, J.~D. Edelstein, J.~Maldacena, and A.~Zhiboedov, {\slshape
  {Causality constraints on corrections to the graviton three-point coupling},}
  \href{http://dx.doi.org/10.1007/JHEP02(2016)020}{{\em JHEP} {\bfseries 02}
  (2016) 020}, \href{http://arxiv.org/abs/1407.5597}{{
  arXiv:1407.5597~[hep-th]}}.

\bibitem{Okubo:1963fa}
S.~Okubo, {\slshape {Phi meson and unitary symmetry model},}
  \href{http://dx.doi.org/10.1016/S0375-9601(63)92548-9}{{\em Phys. Lett.}
  {\bfseries 5} (1963) 165--168}.

\bibitem{Zweig:1964jf}
G.~Zweig, {\em {An SU(3) model for strong interaction symmetry and its
  breaking. Version 2}},
  \href{http://dx.doi.org/10.17181/CERN-TH-412}{pp.~22--101}.
\newblock Hadronic Press, Nonantum, MA, 2, 1964.

\bibitem{Iizuka:1966fk}
J.~Iizuka, {\slshape {Systematics and phenomenology of meson family},}
  \href{http://dx.doi.org/10.1143/PTPS.37.21}{{\em Prog. Theor. Phys. Suppl.}
  {\bfseries 37} (1966) 21--34}.

\bibitem{Correia:2020xtr}
M.~Correia, A.~Sever, and A.~Zhiboedov, {\slshape {An analytical toolkit for
  the S-matrix bootstrap},}
  \href{http://dx.doi.org/10.1007/JHEP03(2021)013}{{\em JHEP} {\bfseries 03}
  (2021) 013}, \href{http://arxiv.org/abs/2006.08221}{{
  arXiv:2006.08221~[hep-th]}}.

\bibitem{Buric:2023ykg}
I.~Buric, F.~Russo, and A.~Vichi, {\slshape {Spinning Partial Waves for
  Scattering Amplitudes in $d$ Dimensions},}
  \href{http://arxiv.org/abs/2305.18523}{{ arXiv:2305.18523~[hep-th]}}.

\bibitem{sdpb}
D.~Simmons-Duffin, {\slshape {A Semidefinite Program Solver for the Conformal
  Bootstrap},} \href{http://dx.doi.org/10.1007/JHEP06(2015)174}{{\em JHEP}
  {\bfseries 06} (2015) 174}, \href{http://arxiv.org/abs/1502.02033}{{
  arXiv:1502.02033~[hep-th]}}.

\bibitem{Caracciolo:2009bx}
F.~Caracciolo and V.~S. Rychkov, {\slshape {Rigorous Limits on the Interaction
  Strength in Quantum Field Theory},}
  \href{http://dx.doi.org/10.1103/PhysRevD.81.085037}{{\em Phys. Rev. D}
  {\bfseries 81} (2010) 085037}, \href{http://arxiv.org/abs/0912.2726}{{
  arXiv:0912.2726~[hep-th]}}.

\bibitem{Caron-Huot:2021rmr}
S.~Caron-Huot, D.~Mazac, L.~Rastelli, and D.~Simmons-Duffin, {\slshape {Sharp
  boundaries for the Swampland},}
  \href{http://dx.doi.org/10.1007/jhep07(2021)110}{{\em JHEP} {\bfseries 07}
  (2021) 110}, \href{http://arxiv.org/abs/2102.08951}{{
  arXiv:2102.08951~[hep-th]}}.

\bibitem{Caron-Huot:2022ugt}
S.~Caron-Huot, Y.-Z. Li, J.~Parra-Martinez, and D.~Simmons-Duffin, {\slshape
  {Causality constraints on corrections to Einstein gravity},}
  \href{http://dx.doi.org/10.1007/JHEP05(2023)122}{{\em JHEP} {\bfseries 05}
  (2023) 122}, \href{http://arxiv.org/abs/2201.06602}{{
  arXiv:2201.06602~[hep-th]}}.

\bibitem{Henriksson:2021ymi}
J.~Henriksson, B.~McPeak, F.~Russo, and A.~Vichi, {\slshape {Rigorous bounds on
  light-by-light scattering},}
  \href{http://dx.doi.org/10.1007/JHEP06(2022)158}{{\em JHEP} {\bfseries 06}
  (2022) 158}, \href{http://arxiv.org/abs/2107.13009}{{
  arXiv:2107.13009~[hep-th]}}.

\bibitem{McPeak:2023wmq}
B.~McPeak, M.~Venuti, and A.~Vichi, {\slshape {Adding subtractions: comparing
  the impact of different Regge behaviors},}
  \href{http://arxiv.org/abs/2310.06888}{{ arXiv:2310.06888~[hep-th]}}.

\bibitem{Acanfora:2023axz}
F.~Acanfora, A.~Guerrieri, K.~H\"aring, and D.~Karateev, {\slshape {Bounds on
  scattering of neutral Goldstones},} \href{http://arxiv.org/abs/2310.06027}{{
  arXiv:2310.06027~[hep-th]}}.

\bibitem{Komargodski:2016auf}
Z.~Komargodski and D.~Simmons-Duffin, {\slshape {The Random-Bond Ising Model in
  2.01 and 3 Dimensions},}
  \href{http://dx.doi.org/10.1088/1751-8121/aa6087}{{\em J. Phys. A} {\bfseries
  50} (2017) 154001}, \href{http://arxiv.org/abs/1603.04444}{{
  arXiv:1603.04444~[hep-th]}}.

\bibitem{Simmons-Duffin:2016wlq}
D.~Simmons-Duffin, {\slshape {The Lightcone Bootstrap and the Spectrum of the
  3d Ising CFT},} \href{http://dx.doi.org/10.1007/JHEP03(2017)086}{{\em JHEP}
  {\bfseries 03} (2017) 086}, \href{http://arxiv.org/abs/1612.08471}{{
  arXiv:1612.08471~[hep-th]}}.

\bibitem{Haring:2023zwu}
K.~H\"aring and A.~Zhiboedov, {\slshape {The Stringy S-matrix Bootstrap:
  Maximal Spin and Superpolynomial Softness},}
  \href{http://arxiv.org/abs/2311.13631}{{ arXiv:2311.13631~[hep-th]}}.

\bibitem{Katz:2005ir}
E.~Katz, A.~Lewandowski, and M.~D. Schwartz, {\slshape {Tensor mesons in
  AdS/QCD},} \href{http://dx.doi.org/10.1103/PhysRevD.74.086004}{{\em Phys.
  Rev. D} {\bfseries 74} (2006) 086004},
  \href{http://arxiv.org/abs/hep-ph/0510388}{{ arXiv:hep-ph/0510388}}.

\bibitem{sakurai}
J.~Sakurai, {\em Currents and mesons}.
\newblock C9B43 K9.

\bibitem{rhos}
J.~Albert, J.~Henriksson, L.~Rastelli, and A.~Vichi. In preparation.

\bibitem{Caron-Huot:2016icg}
S.~Caron-Huot, Z.~Komargodski, A.~Sever, and A.~Zhiboedov, {\slshape {Strings
  from Massive Higher Spins: The Asymptotic Uniqueness of the Veneziano
  Amplitude},} \href{http://dx.doi.org/10.1007/JHEP10(2017)026}{{\em JHEP}
  {\bfseries 10} (2017) 026}, \href{http://arxiv.org/abs/1607.04253}{{
  arXiv:1607.04253~[hep-th]}}.

\bibitem{Veneziano:2017cks}
G.~Veneziano, S.~Yankielowicz, and E.~Onofri, {\slshape {A model for pion-pion
  scattering in large-N QCD},}
  \href{http://dx.doi.org/10.1007/JHEP04(2017)151}{{\em JHEP} {\bfseries 04}
  (2017) 151}, \href{http://arxiv.org/abs/1701.06315}{{
  arXiv:1701.06315~[hep-th]}}.

\bibitem{Caron-Huot:2022jli}
S.~Caron-Huot, Y.-Z. Li, J.~Parra-Martinez, and D.~Simmons-Duffin, {\slshape
  {Graviton partial waves and causality in higher dimensions},}
  \href{http://dx.doi.org/10.1103/PhysRevD.108.026007}{{\em Phys. Rev. D}
  {\bfseries 108} (2023) 026007}, \href{http://arxiv.org/abs/2205.01495}{{
  arXiv:2205.01495~[hep-th]}}.

\bibitem{Bianchi:2020cfc}
M.~Bianchi, D.~Consoli, and P.~Di~Vecchia, {\slshape {On the N-pion extension
  of the Lovelace-Shapiro model},}
  \href{http://dx.doi.org/10.1007/JHEP03(2021)119}{{\em JHEP} {\bfseries 03}
  (2021) 119}, \href{http://arxiv.org/abs/2002.05419}{{
  arXiv:2002.05419~[hep-th]}}.

\end{thebibliography}\endgroup
\end{document}